\documentclass[pre,twocolumn,longbibliography,amsmath,amssymb,floatfix]{revtex4-1}

\usepackage{graphicx}
\usepackage{dcolumn}
\usepackage{bm}
\usepackage[usenames,dvipsnames]{xcolor}
\usepackage{mciteplus}
\usepackage{braket}
\usepackage{ulem}

\newcommand*\dif{\mathop{}\!\mathrm{d}}

\begin{document}

\title{Dynamics of partially thermalized solutions of the Burgers
  equation }
\thanks{Postprint version of the manuscript published in Phys. Rev. Fluids {\bf
3}, 014603 (2018)}
\author{Patricio Clark di Leoni$^{1,2}$\email{clark@df.uba.ar},
  Pablo D.~Mininni$^1$\email{mininni@df.uba.ar}, \&
  Marc E.~Brachet$^3$\email{brachet@physique.ens.fr}}
\affiliation{$^1$Departamento de F\'\i sica, Facultad de Ciencias
Exactas y Naturales, Universidad de Buenos Aires and IFIBA,
CONICET, Ciudad Universitaria, 1428 Buenos Aires, Argentina.
  \\
  $^2$ Department of Physics and INFN, University of Rome ``Tor
  Vergata'', Via della Ricerca Scientifica 1, 00133, Rome, Italy.
  \\
  $^3$ Laboratoire de Physique Statistique, \'{E}cole Normale
  Sup{\'e}rieure, PSL Research University; UPMC Univ Paris 06,
  Sorbonne Universit\'{e}s; Universit\'{e} Paris Diderot, Sorbonne
  Paris-Cit\'{e}; CNRS; 24 Rue Lhomond, 75005 Paris, France.}
\date{\today}

\begin{abstract}
    The spectrally truncated, or finite dimensional, versions of
    several equations of inviscid flows display transient solutions
    which match their viscous counterparts, but which 
    eventually lead to thermalized states in which energy is in
    equipartition between all modes. Recent advances in the study of
    the Burgers equation show that the thermalization process is
    triggered after the formation of sharp localized structures within
    the flow called ``tygers''. We show that the process of
    thermalization first takes place in well defined subdomains,
    before engulfing the whole space. Using spatio-temporal analysis
    on data from numerical simulations, we study propagation of
    tygers and find that they move at a well defined mean speed that
    can be obtained from energy conservation arguments.
\end{abstract}

\maketitle

\section{Introduction} 

The formulation of a proper microscopic theory of turbulence has
remained a major challenge in statistical physics \cite{Kraichnan89}.
In spite of this, classic Gibbs ensembles have provided significant
insights when used to predict the equilibrium of ideal flows in which
only a finite number of spatial modes are allowed 
\cite{Lee52,Kraichnan67,Ting86}. For the case of the spectrally
truncated three-dimensional Euler equation, energy is then
equipartitioned between all modes in a thermalized equilibrium
state. This results in an energy spectrum that goes like $\sim
k^2$, which differs greatly from the $\sim k^{-5/3}$ Kolmogorov
spectrum that is in agreement with experiments and observations. The
main hurdle is that macroscopic hydrodynamics is essentially
dissipative, and so conservative statistical formulations cannot
capture its essence.

Nonetheless, statistical equilibra of spectrally truncated systems have
played an important role in turbulence theory. The main reason for this
is that they give a proxy for the direction of the energy cascade in the
forced and dissipative case \cite{Kraichnan67}, and they allow
identification of attractors in freely decaying dissipative cases
\cite{Ting86}. As  an example of the former, they have led to the
prediction of the inverse energy cascade in two-dimensional turbulence
\cite{Kraichnan67}, while an example of the latter is
magnetohydrodynamics, where the existence of multiple quadratic
invariants results in several long-time possible solutions which were
identified using statistical equilibria \cite{Ting86}.

Recently the interest on Gibbs ensembles in turbulence was renewed, as
it was also found that the transient as the ideal truncated system
reaches the equilibrium can mimic forced and dissipative systems. It was
originally suggested by Kraichnan and Chen \cite{Kraichnan89} that
truncated conservative systems can behave as dissipative when
considering only the spatial modes which have not thermalized. The idea
behind this is that high wave number thermalized modes can act as an
energy sink for the low wave number modes, which will behave as in a
normal turbulent flow. This has been put on firm grounds by computing
eddy viscosity caused by thermalized modes and confirmed numerically in
high resolution simulations of the Euler equation in \cite{Cichowlas05}.
In the simulations, energy was initially concentrated at low wave
numbers and was let to cascade to larger ones, and a long transient
following the previous description was observed before the system
reached full thermalization. The results were extended to helical
hydrodynamic flows \cite{Krstulovic09b}, magnetohydrodynamics
\cite{Krstulovic11}, compressible flows \cite{Krstulovic09}, quantum
turbulence \cite{Krstulovic11b,Zhu16}, gyrokinetic plasma systems
\cite{Zhu10}, the dyamo problem \cite{Dmitruk14,Prasath14}, and also
to study the decay of quasi-geostrophic turbulence
\cite{Teitelbaum12}.

While systems such as the spectrally truncated Euler equation are
known to thermalize and to have a viscous-like transient, not much is
known about how thermalization begins and evolves, or about how 
the limit of the truncation wave number going to infinity behaves.  The
recent discovery of a phenomenon dubbed as ``tygers'' 
\cite{Ray11} in studies of the two-dimensional Euler and of the
Burgers equations has opened a new path to tackle these problems. The
inviscid Burgers equation is a non-linear PDE known to develop shocks,
for which energy-preserving truncations do not converge (in a weak
sense) to the inviscid limit \cite{Lax79,Goodman88,Hou91}, and whose
spectrally truncated version thermalizes in finite time
\cite{Majda00}. The curious fact is that the first ``spurious''
effects of thermalization in physical space do not occur near the
shock, but away from it. Sharp localized structures, the so-called
tygers, are formed.  After collapsing, thermalization starts to take
place near the location of the tyger, eventually expanding to the
whole domain.  The mechanism behind the formation of a tyger has been
identified as a resonant interaction between fluid particles and
truncation noise \cite{Ray11}. Further studies \cite{Venkataraman17}
have determined the time between the appearance of a tyger and the
onset of thermalization, and its scaling with the truncation
wave number. Recent studies have linked the appearance of tygers to a
period-doubling bifurcation and loss of stability of the truncated wave
solutions \cite{Feng17}.

\begin{figure*}
    \centering
    \includegraphics[width=0.3\textwidth]{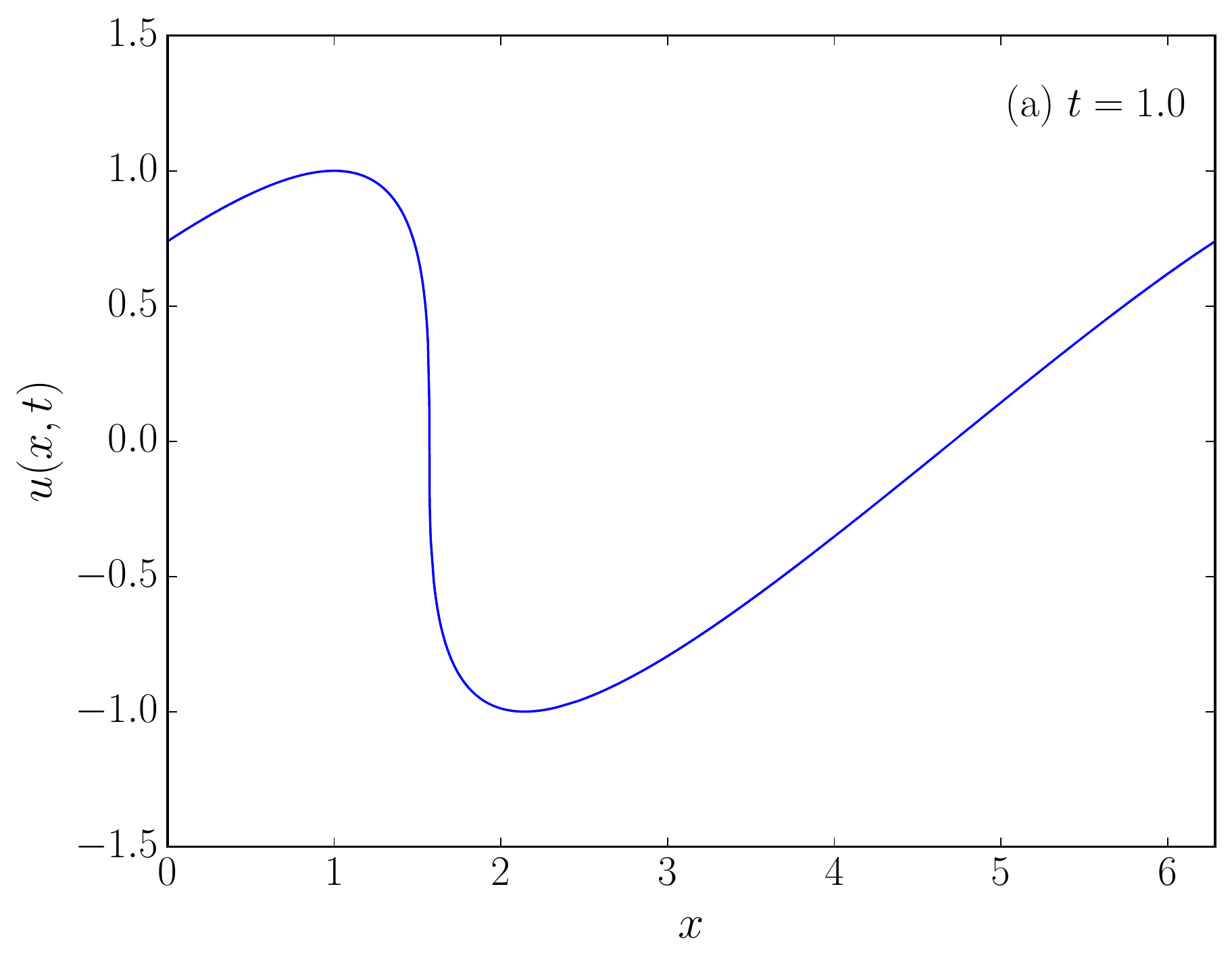}
    \includegraphics[width=0.3\textwidth]{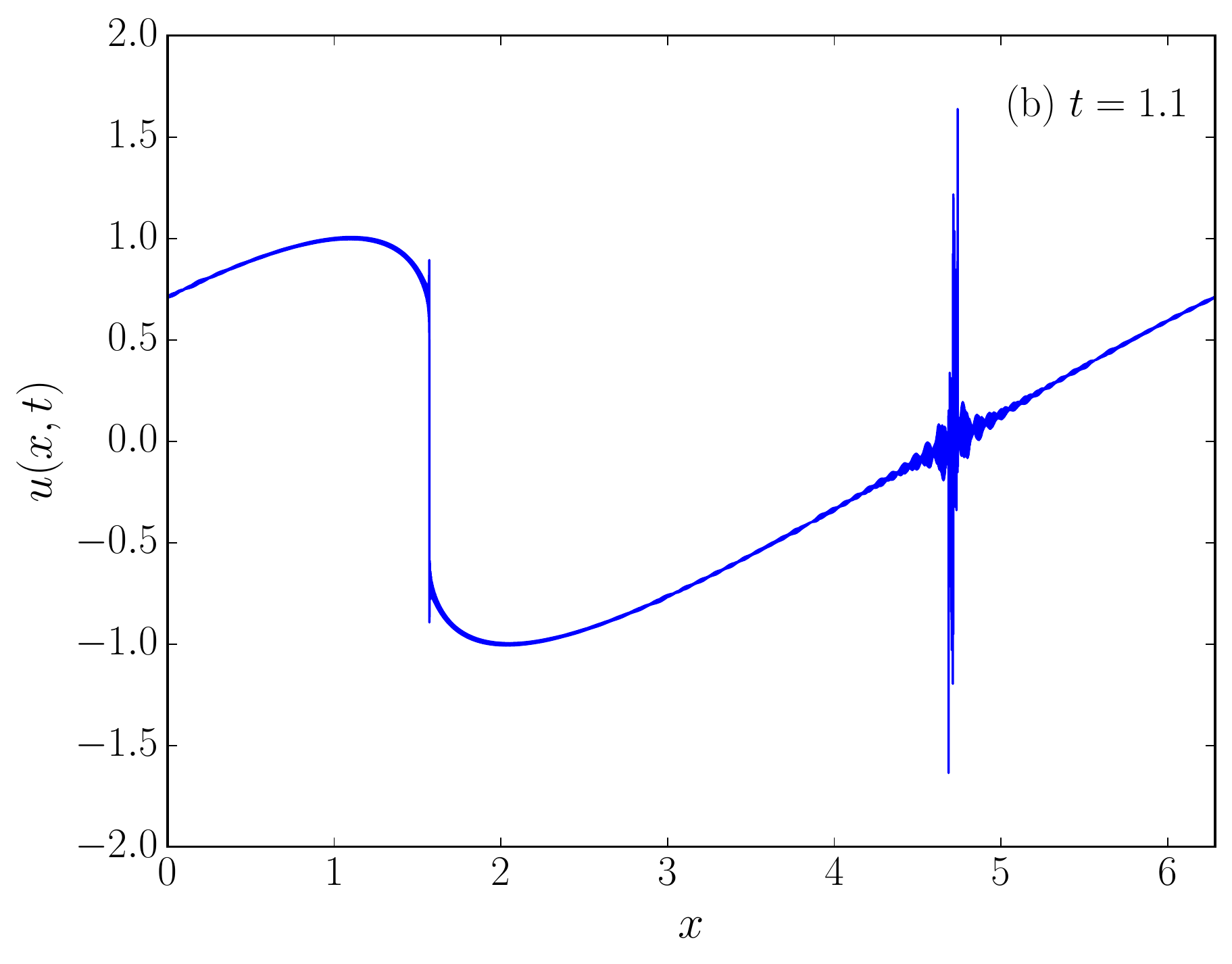}
    \includegraphics[width=0.3\textwidth]{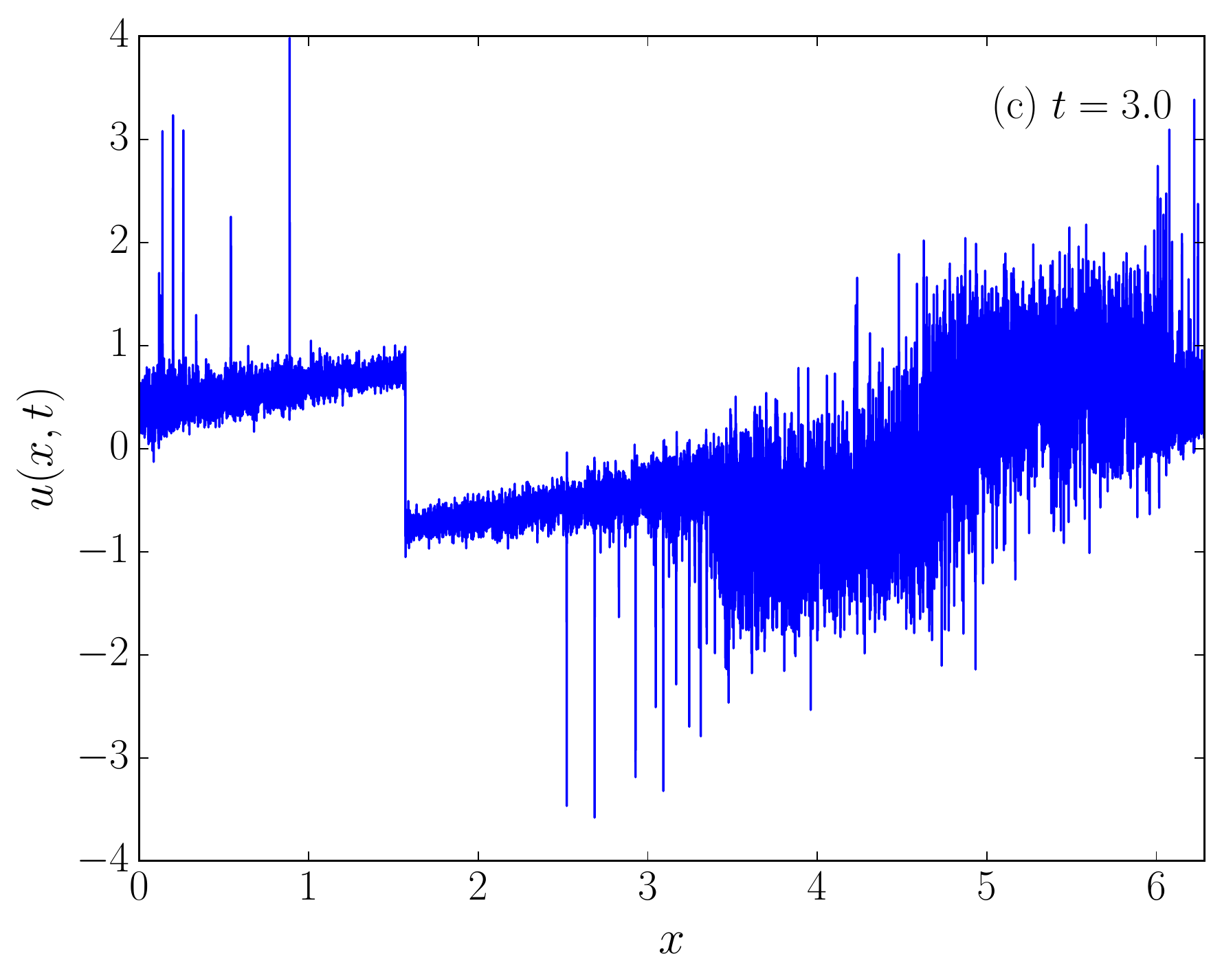}
    \includegraphics[width=0.3\textwidth]{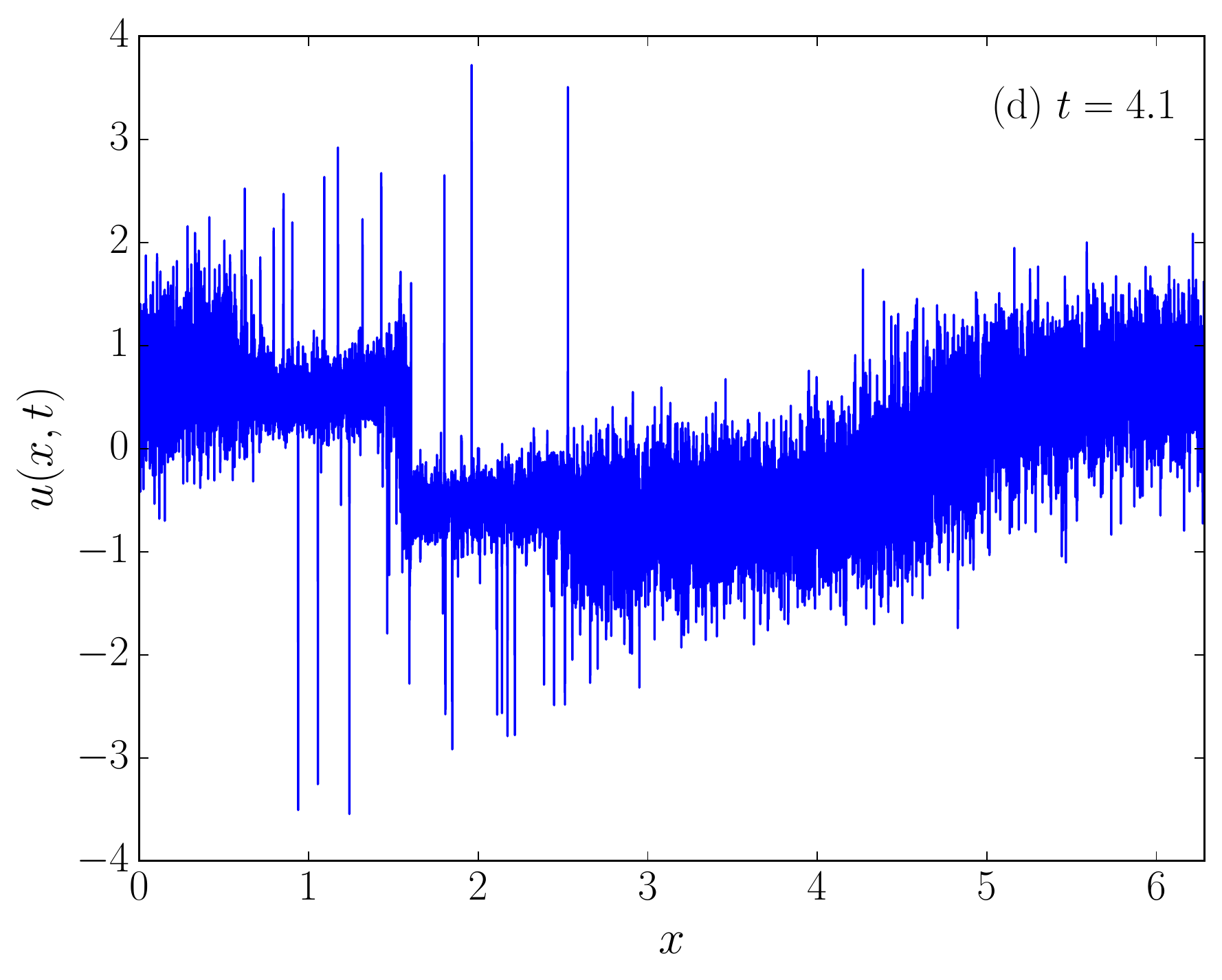}
    \includegraphics[width=0.3\textwidth]{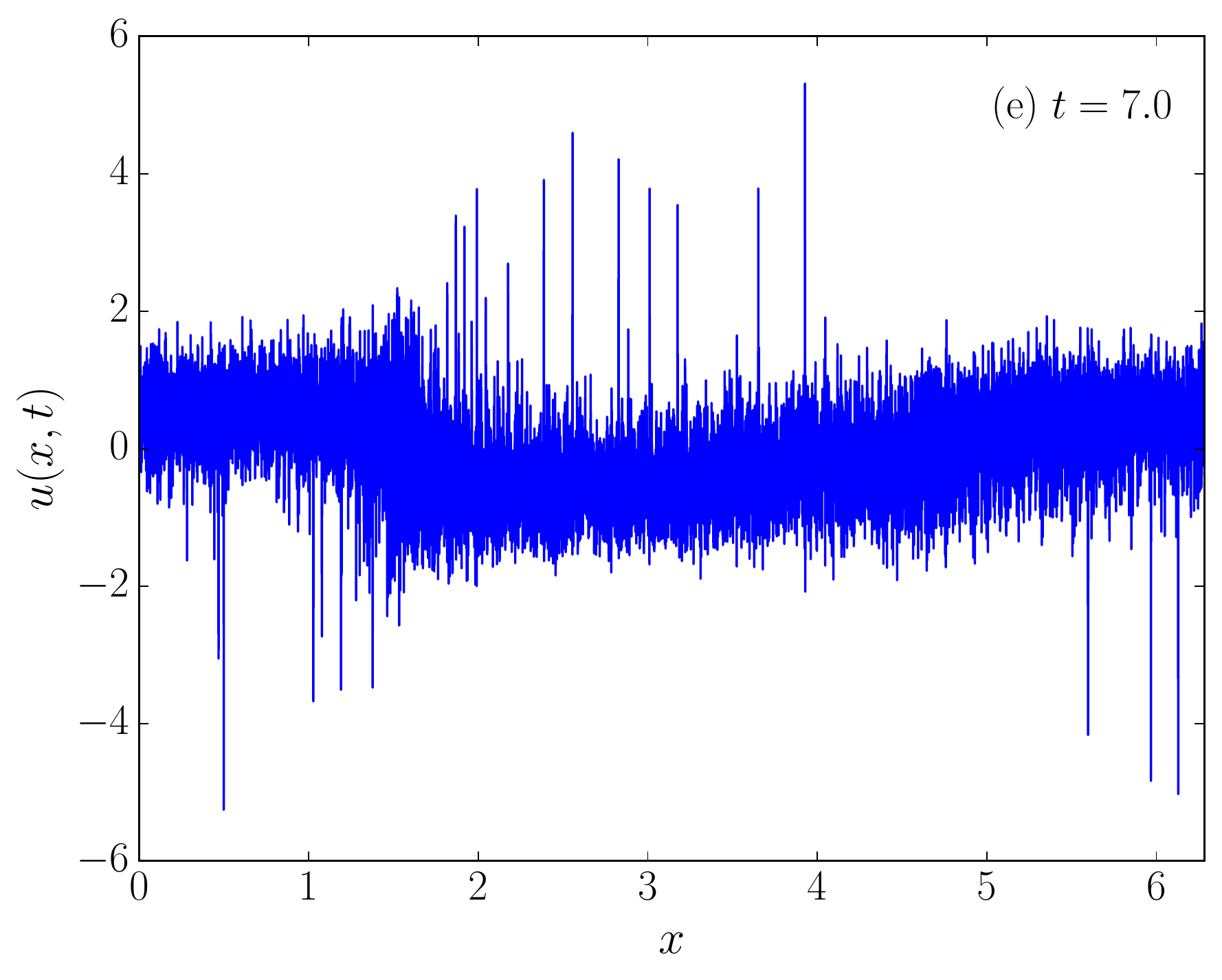}
    \includegraphics[width=0.3\textwidth]{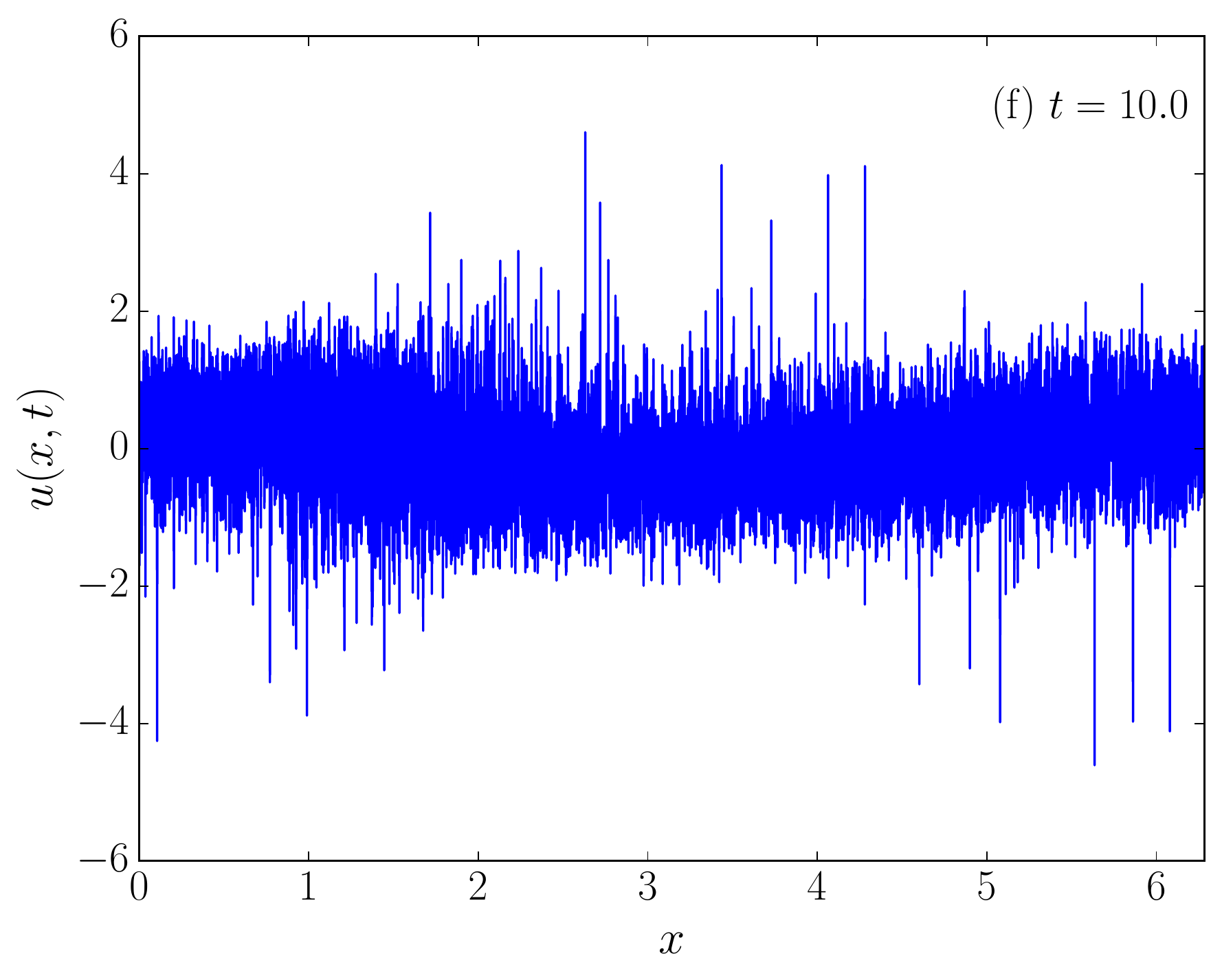}
    \caption{({\it Color online}) Evolution of $u(x,t)$ at different
      times, from (a) $t=1$ to (f) $t=10$, in a a simulation with 
      $k_G = 5461$ and a single-mode initial condition given by
      Eq.~\eqref{onemode}. The shock is formed at $t=1$, and the tyger
      has developed and started to collapse at $t=1.1$. Afterwards
      each front of the tyger advances, swallowing up the rest of the
      solution. (a) $t=1.0$, (b) $t=1.1$, (c) $t=3.0$, (d) $t=4.1$, (e)
      $t=7.0$, and (f) $10.0$.}
    \label{tempevol}
\end{figure*}

The importance of tygers and the thermalization mechanism is multiple.
The Burgers equation has been used as a toy model of turbulence which
also displays shocks (see, for example, \cite{Bec07,Buzzicotti16}
and references therein). As such, the study of tygers gives information
on dynamical processes in the system, and on how modes interact
non-linearly. They are also important for numerical methods as they
develop and increase error in regions in which the flow is initially
smooth, and as their control or removal would allow the development of
new methods to integrate equations in the inviscid limit. In
particular, when studying numerically the blow-up problem in fluid
dynamics (i.e., the formation of a singularity in finite time, for
references see \cite{Frisch03,Gibbon08}), attention should be put in
discerning between tygers and actual blow-up effects.

The aim of this paper is to study the evolution of thermalization
itself, and how the first traces of thermalization end up engulfing the
whole domain. We do this via numerical simulation of the
one-dimensional Burgers equation at different truncation
wave numbers. It is found that the system first partially thermalizes
inside different spatial subdomains, defined by the initial positions
of the tygers and shocks. The boundary of the thermalized component
then propagates as the system reaches equilibrium with a well defined
mean velocity. So, while at long times full stochastic behavior is
obtained \cite{Majda00}, at intermediate times deterministic and
stochastic behaviors are mixed.

\section{The Burgers equation}

The inviscid one-dimensional Burgers equation, in conservation form,
reads
\begin{equation}
    \frac{\partial u}{\partial t} +  \frac{\partial
        }{\partial x} \left( \frac{1}{2} u^2 \right)= 0,
    \label{burgers}
\end{equation}
where $u$ is the velocity field. Under periodic boundary conditions
the solutions can be expanded in wave number space in a Fourier series
of the form
\begin{equation}
    u(x,t) = \sum^{\infty}_{k=-\infty} \hat{u}_k(t) e^{ikx},
\end{equation}
where $\hat{u}_k(t)$ are the coefficients of the expansion. As we are
interested in working with the spectrally truncated version of the
equations, we define the Galerkin projector
\begin{equation}
    P_{k_G} [u(x,t)] = \sum_{|k|\leq k_G} \hat{u}_k(t) e^{ikx} ,
    \label{truncation}
\end{equation}
which is just a low-pass filter that sets all Fourier modes with wave
number $|k|>k_G$ to zero, and where $k_G$ is the truncation wave
number. The spectrally truncated Burgers equation then reads
\begin{equation}
    \frac{\partial}{\partial t} P_{k_G} (u) + \frac{1}{2} 
    \frac{\partial}{\partial x} P_{k_G} \left(u^2 \right) = 0.
    \label{galerkin}
\end{equation}
This equation conserves linear momentum $\sum u$ and energy 
$\sum u^2$. It is a well known fact that the continuum
(untruncated) Eq.~\eqref{burgers} produces a shock in finite time. 
For the truncated Eq.~\eqref{galerkin} it has been shown that a
resonant interaction between the fluid particles and the truncation
noise causes the formation of sharp localized structures, the
so-called tygers, in regions in which the flow is smooth
\cite{Ray11}. After formation, tygers then collapse and give rise to
thermalization. The timescale under which the collapse happens scales
as $\sim k^{-4/9}_G$ \cite{Venkataraman17}. Then, for sufficiently
long times, the system reaches full thermalization and all its
properties can be predicted using the Gibbs canonical ensemble with
partition function 
\begin{equation}
    Z = C_\beta e^{-\beta \sum^{k_G}_{k=1} |\hat{u}_k|^2},
\end{equation}
where $\beta = k_G/\bar{E}$, and where $\bar{E}$ is the mean energy of
the flow \cite{Majda00}. For details on the formation of the tygers
the reader is referred to the studies in
\cite{Ray11,Venkataraman17,Feng17},
and for the thermalized solutions at late times to \cite{Majda00}. In
the following we will be concerned with what happens between the
triggering of thermalization and until the subsequent statistical
equilibrium is reached.

\begin{figure}
    \centering
    \includegraphics[trim={1.6cm .5cm .8cm 0},clip,width=8.5cm]{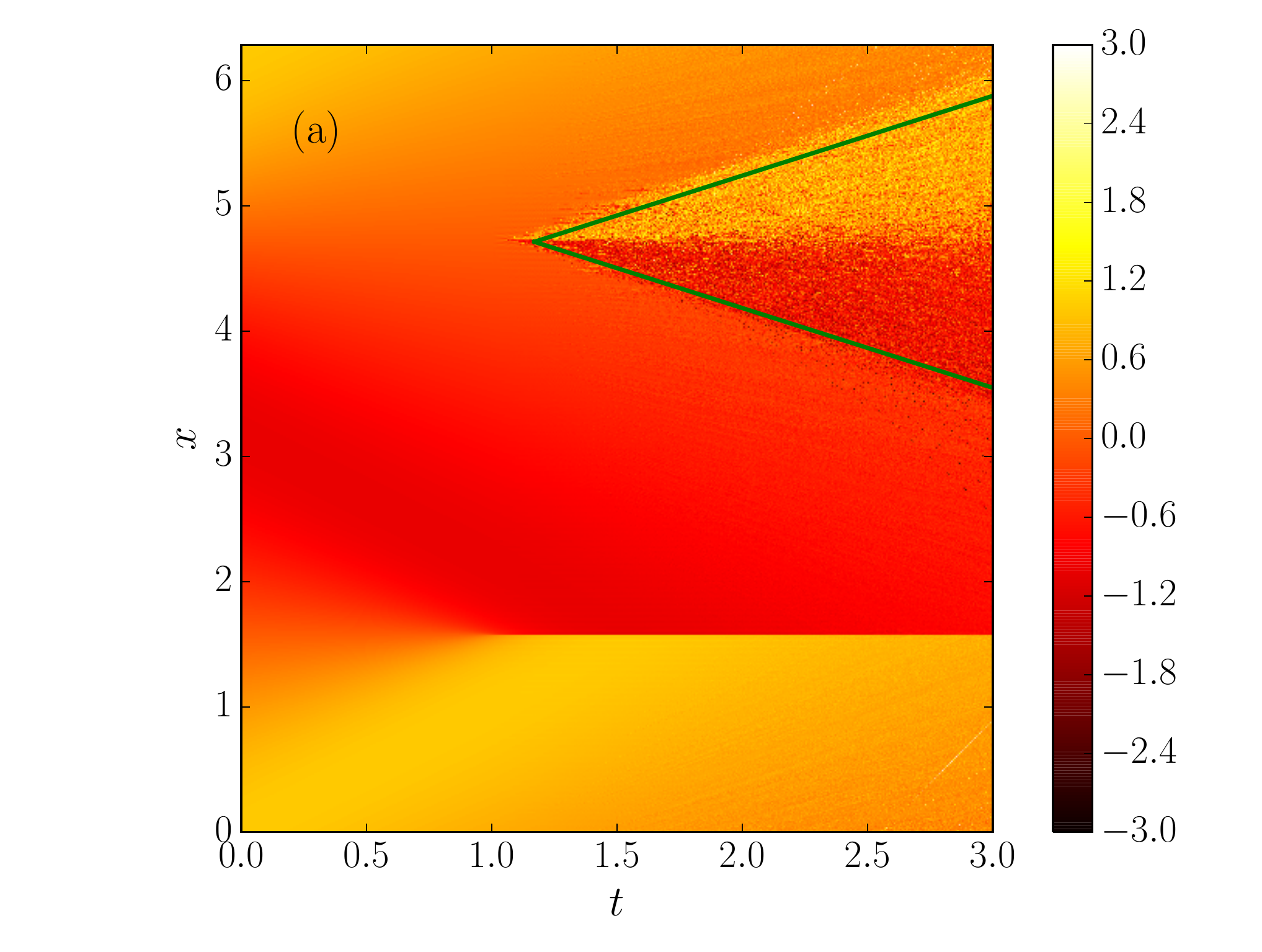}
    \includegraphics[width=8.5cm]{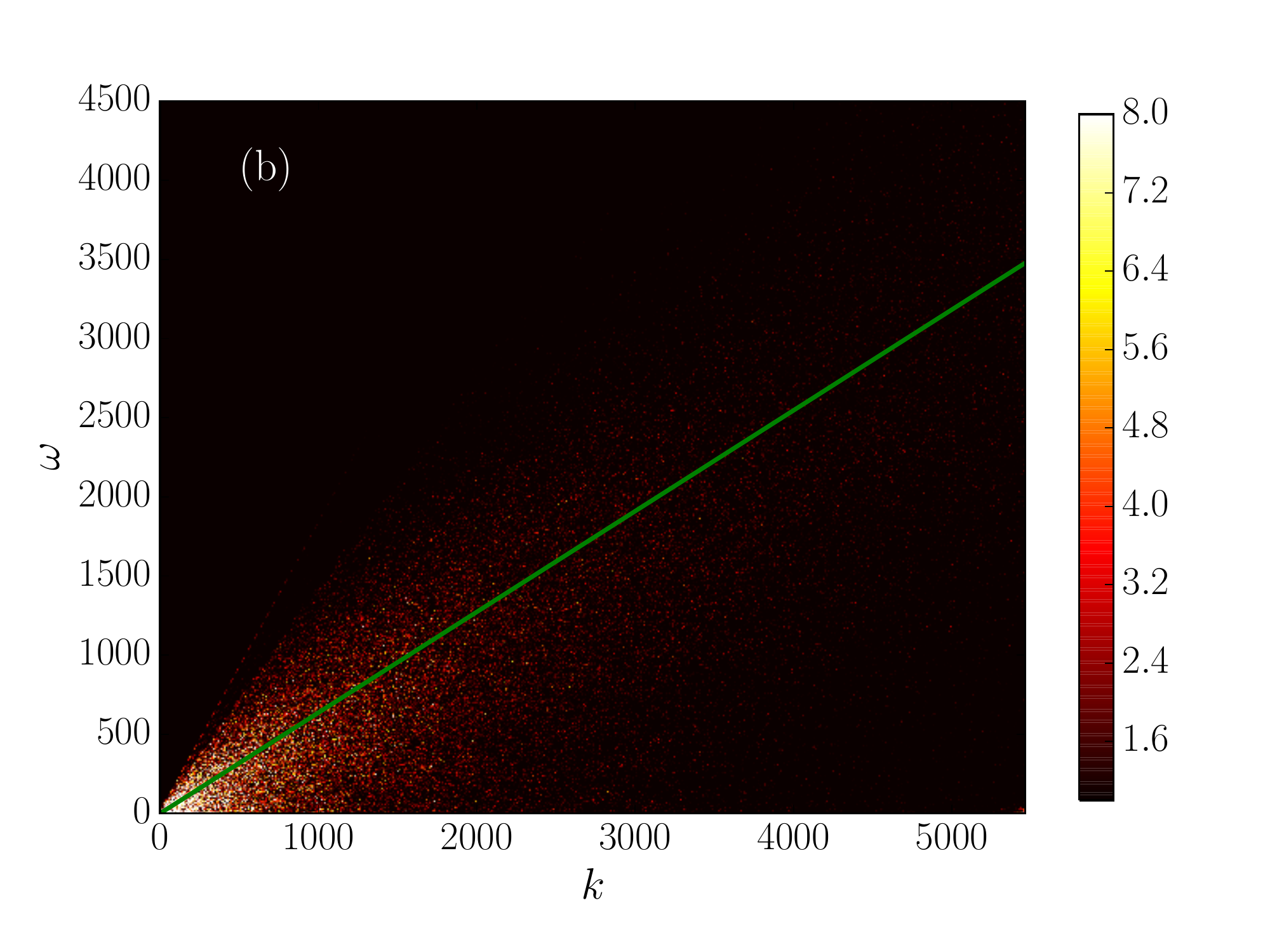}
    \caption{({\it Color online}) Spatiotemporal evolution of $u(x,t)$:
    (a) Evolution in real space as a function of space and time, and (b)
    evolution in Fourier space as a function of frequency and wave
    number. The spreading of the tyger after $t=1.1$ is well
    described by a velocity (marked in both plots with a green solid
    line) equal to $2/\pi$. This is the mean velocity of propagation of
    each thermalized subdomain formed to the right and left of the tyger
    when the tyger appears.}
    \label{sptmp}
\end{figure}

\section{Numerical simulations}

For the purpose of this study Eq.~\eqref{galerkin} was solved
numerically using a pseudospectral method with the $2/3$ rule for
dealiasing, which naturally implements the spectral truncation, and
which conserves the energy. Time integration was done using a
Runge-Kutta method. Different truncation wave numbers were considered,
from $k_G = 341$ to $k_G = 10922$. By virtue of the $2/3$ rule, the
truncation wave number is $k_G=N/3$ where $N$ is the spatial resolution.
Thus, simulations with spatial resolutions from $N=1024$ to $N=32768$
grid points were done. All of these simulations performed have
essentially the same behavior, and none of the results we now present
depend on $k_G$ (at least for the values of $k_G$ we considered, the
limit of $k_G$ going to infinity is highly non-trivial).  Therefore, all
the figures we present in the main text are from a simulation with
$k_G = 5461$ ($N=16384$). A comparison between the different
resolutions is shown in the Appendix.

Two different initial conditions were used; first, a single-mode initial
condition
\begin{equation}
    u_1(x) = \cos(x),
    \label{onemode}
\end{equation}
and then an initial condition with two excited modes
\begin{equation}
    u_2 (x) = \sin(x) + \sin(2 x - 0.741) ,
    \label{twomode}
\end{equation}
which was used before in \cite{Ray11} to study the early time
development of tygers.

\section{Results} 

Under the single-mode initial condition $u_1(x)$ given by
Eq.~\eqref{onemode}, the system develops a shock a $t=1$ (see
Fig.~\ref{tempevol}). Snapshots of $u(x,t)$ at different times,
ranging from $t=1$ up to $t=10$, are also shown in
Fig.~\ref{tempevol}. At $t=1$ only the shock can be seen at $x=\pi/2$,
while at $t=1.1$ the tyger is already present at $x=3\pi/2$. As
reported in previous studies \cite{Ray11,Venkataraman17,Feng17}, the
tyger develops in a region in which the velocity is smooth, and its
position can be predicted from the fact that it appears at the point
of the flow that has positive strain and that travels with the
same velocity as the shock (in the case of this simulation, $u=0$).

At later times in Fig.~\ref{tempevol} the tyger collapses and starts
to thermalize, seen in the figure as the development of wide regions
that look like white noise (although the shock at $x=\pi/2$ and a
linear ramp in the velocity in the rest of the domain can still be
recognized at $t=3$, $4.1$, and $7$). Note these wide regions
propagate to the left and right, respectively with negative and
positive velocity. Starting from the tyger, thermalization creeps
slowly through space until the whole domain is almost fully
thermalized at $t=10$. How these thermalized fronts propagate
and what are their statistical properties are the main focuses of this
work. From visual inspection, it is easy to see that the velocity of 
these fronts fluctuates around a certain value. We will thus study 
the mean and the variance of these fluctuations (which, as
fluctuations are close to Gaussian, are sufficient to characterize
their statistical properties). Averaging operations should then be
always understood as the spatial average in a certain region of real
space, as defined below. 

\begin{figure*}
    \centering
    \includegraphics[width=0.3\textwidth]{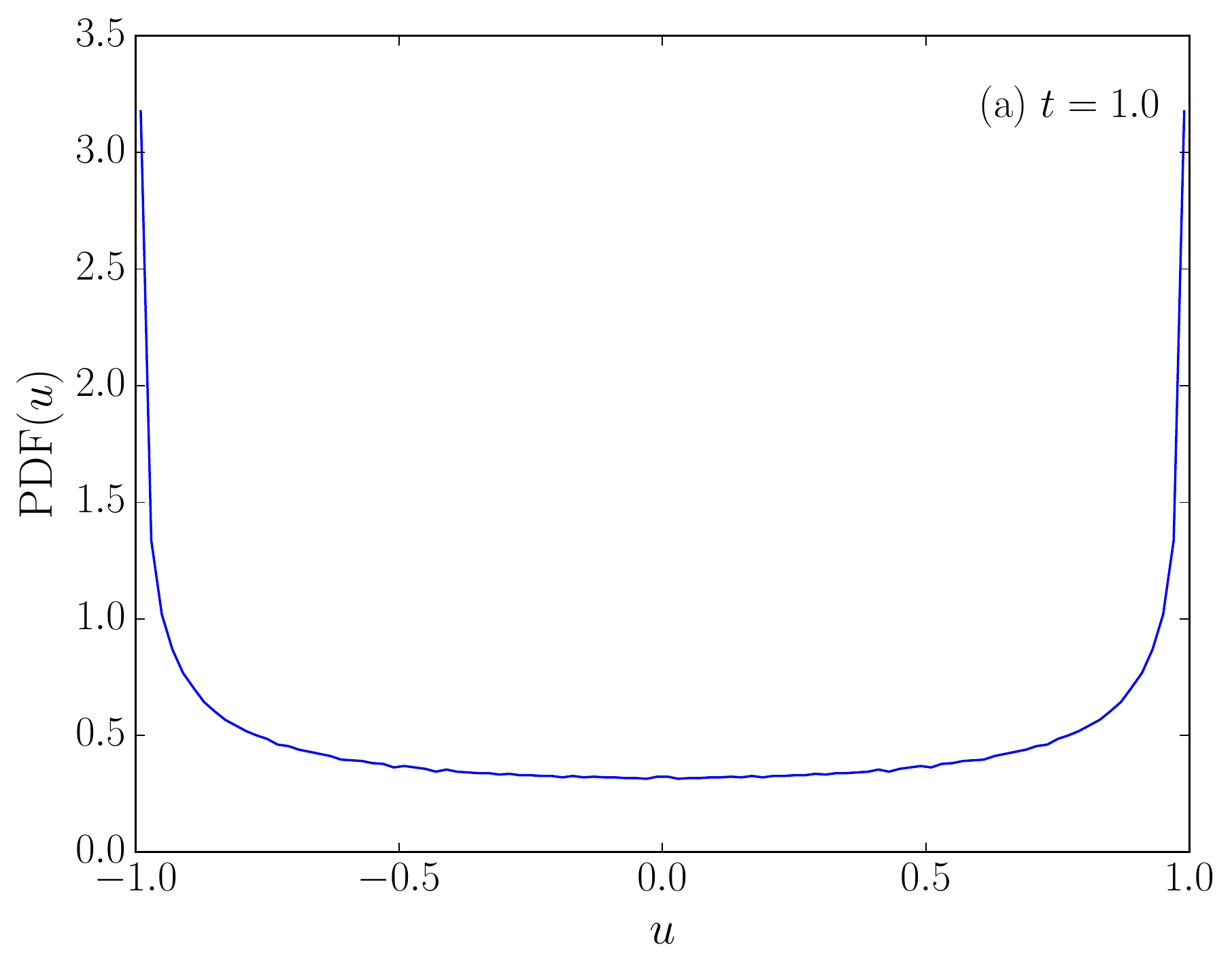}
    \includegraphics[width=0.3\textwidth]{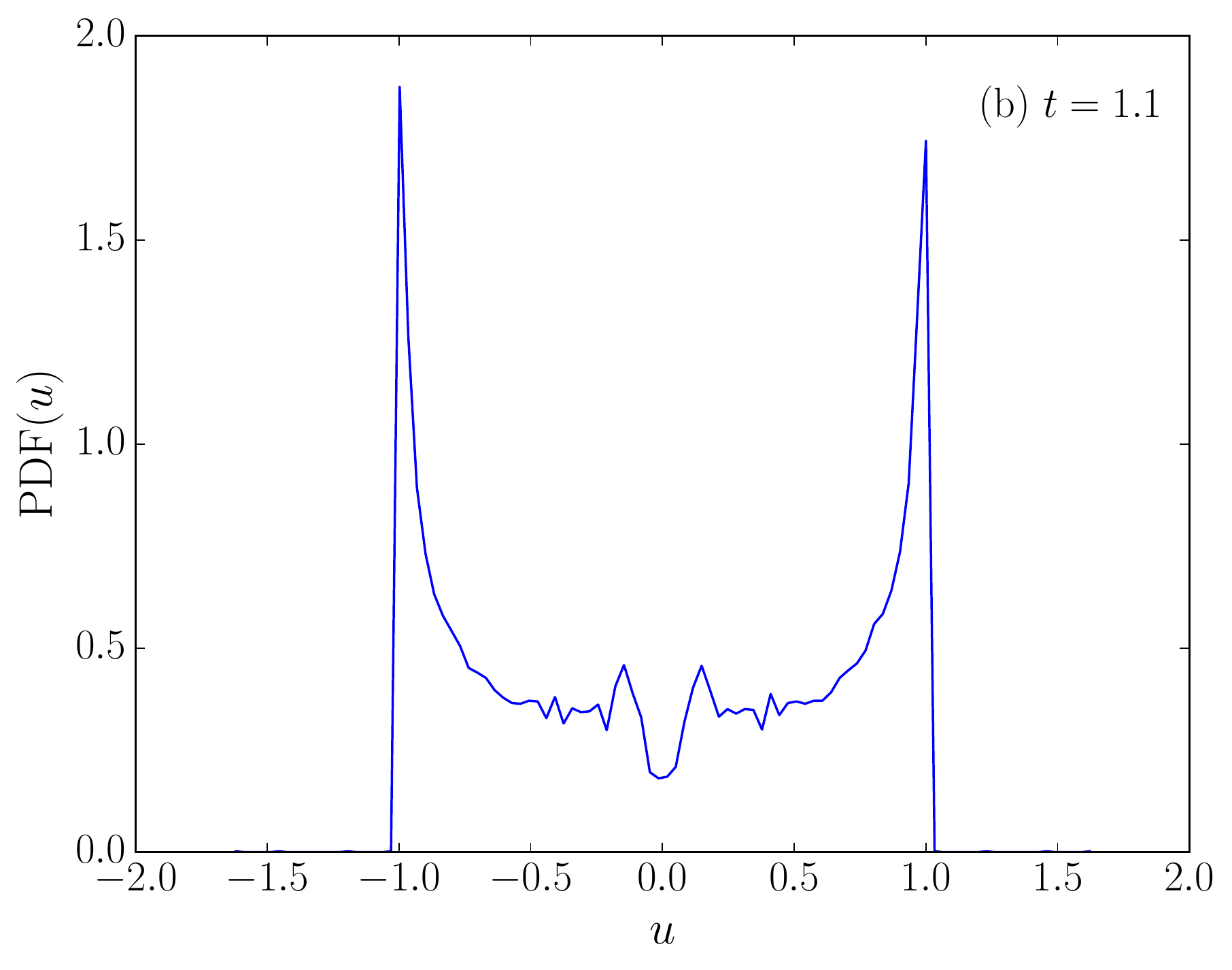}
    \includegraphics[width=0.3\textwidth]{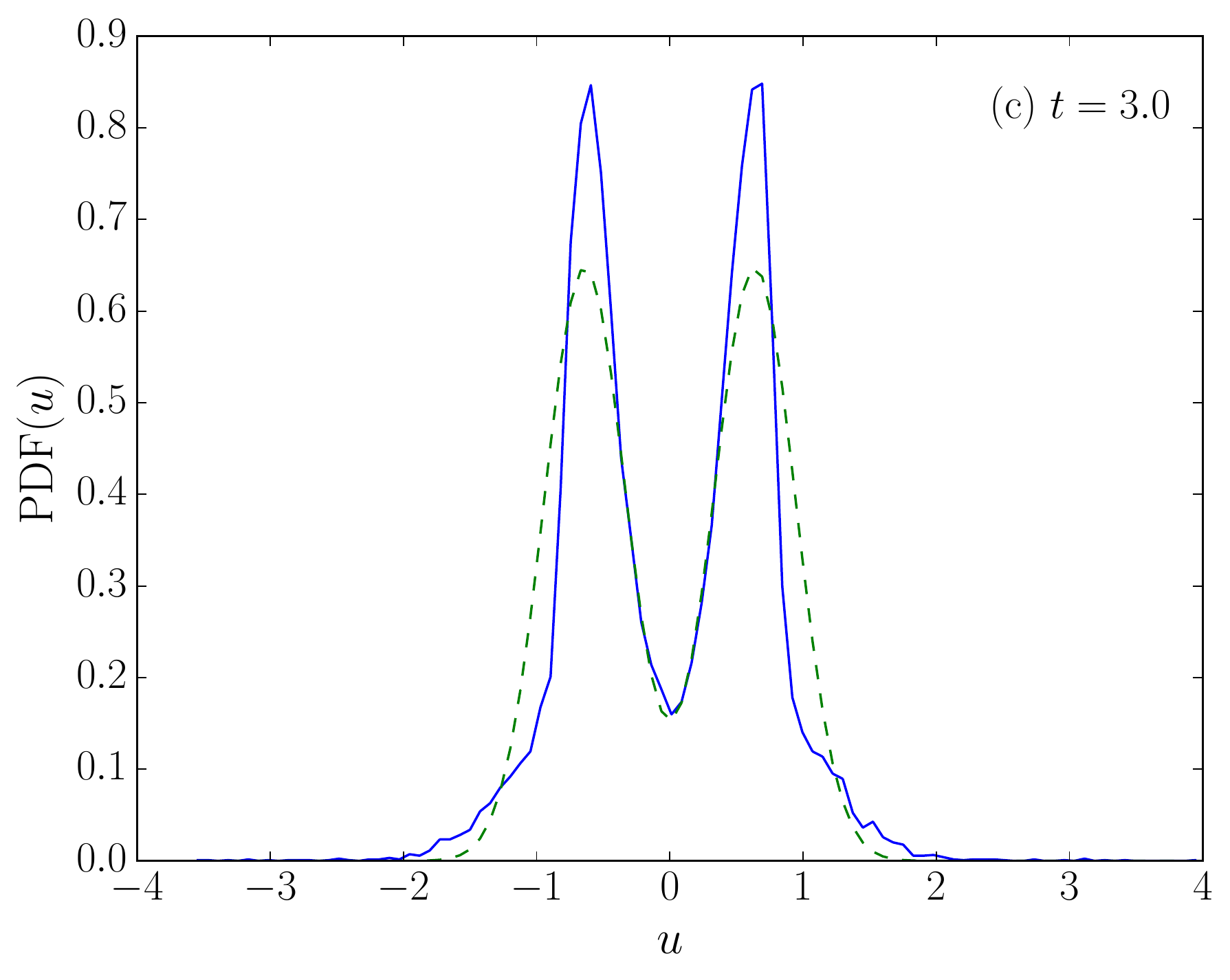}
    \includegraphics[width=0.3\textwidth]{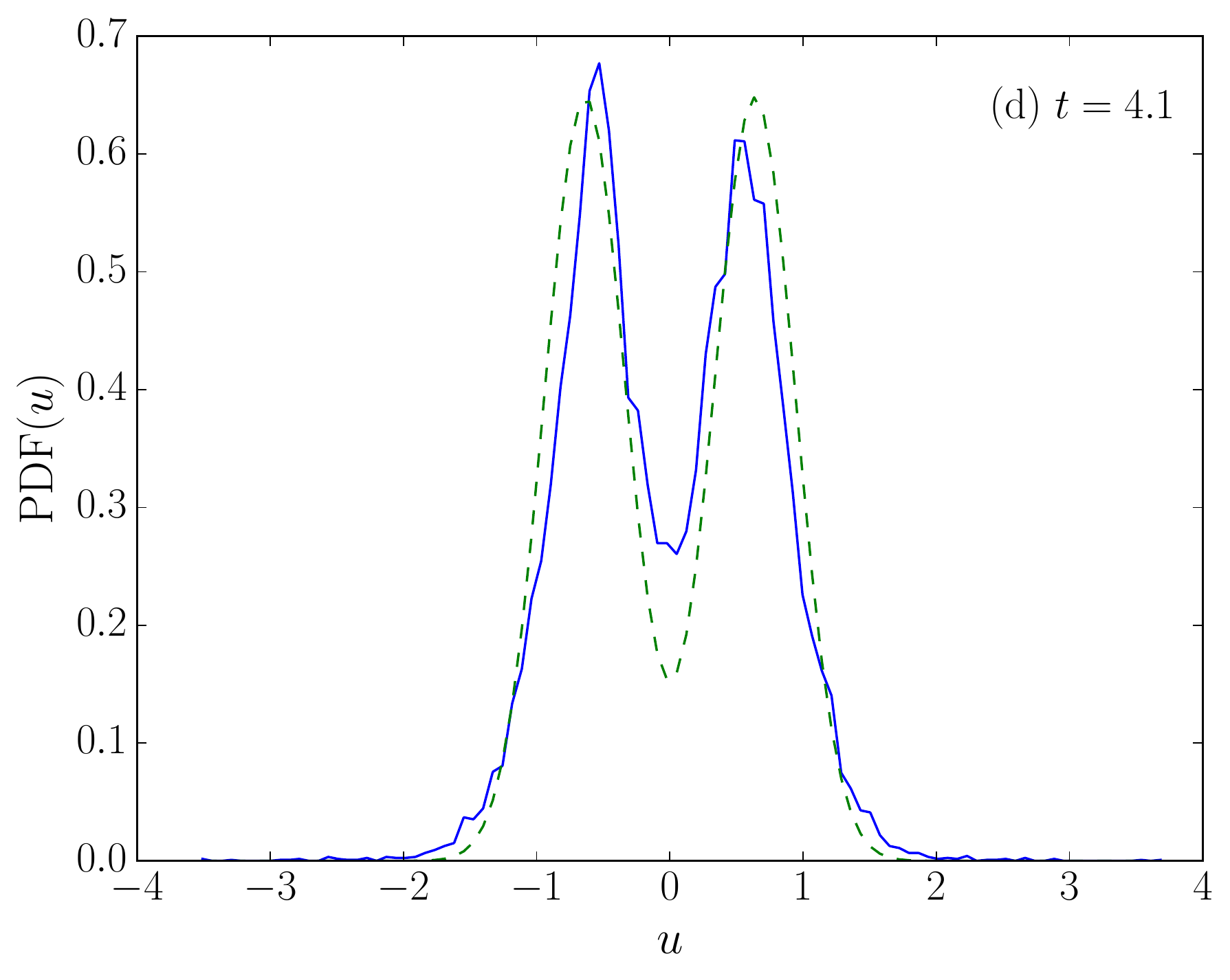}
    \includegraphics[width=0.3\textwidth]{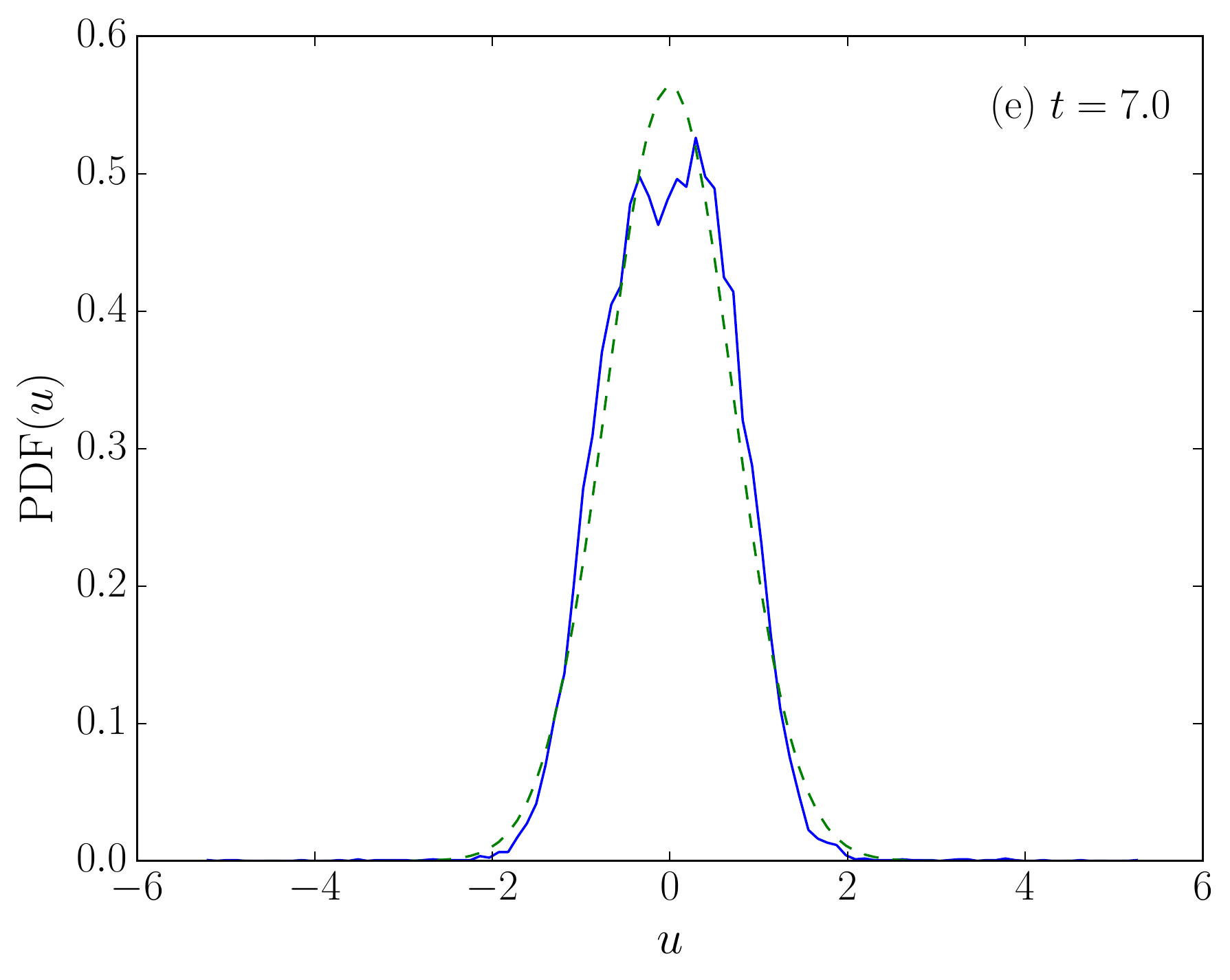}
    \includegraphics[width=0.3\textwidth]{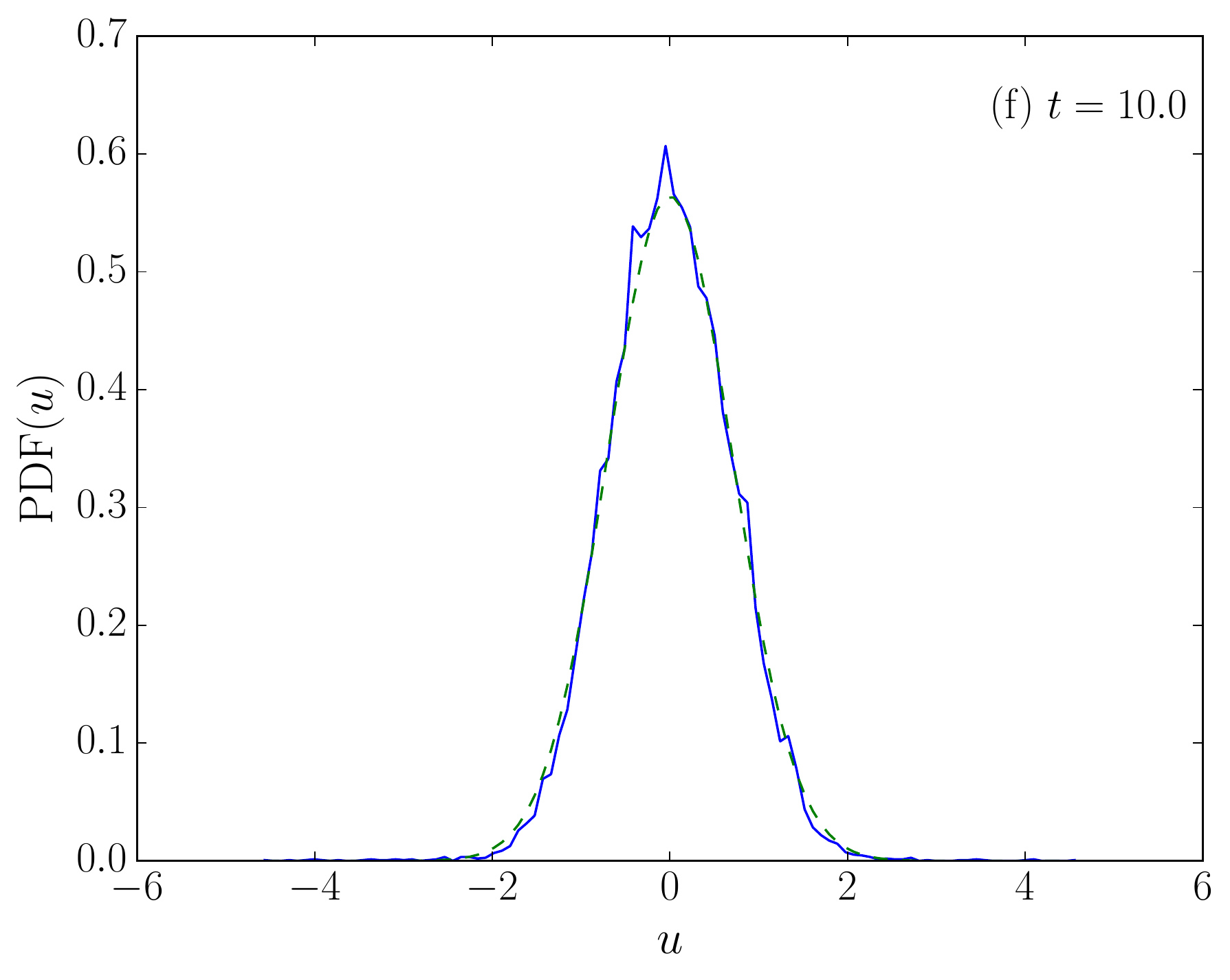}
    \caption{({\it Color online}) Probability density function (PDF) of
    $u(x,t)$ at different times in (blue) solid lines; in (green) dashed
    lines is the proposed PDF for that instant. At the time of shock
    formation (a) the typical PDF of a trigonometric function can be
    observed, then the tyger appears at (b). At intermediate times the
    solution is partially thermalized and a bimodal Gaussian
    distribution gives a good approximation to the data, as seen in (c)
    and (d); the mean and standard deviation of each thermalized mode is
    such that the tyger front matches the statistical properties of the
    solution in the subdomain where it lives. At later times, (e) and
    (f), after the two fronts meet, the system reaches the last stages
    of thermalization and the solution now matches the PDF of white
    noise with statistical properties close to those of the thermal
    equilibrium. (a) $t=1.0$, (b) $t=1.1$, (c) $t=3.0$, (d) $t=4.1$, (e)
    $t=7.0$, and (f) $10.0$.}
    \label{histevol}
\end{figure*}

\begin{figure}[h]
    \centering
    \includegraphics[width=8.5cm]{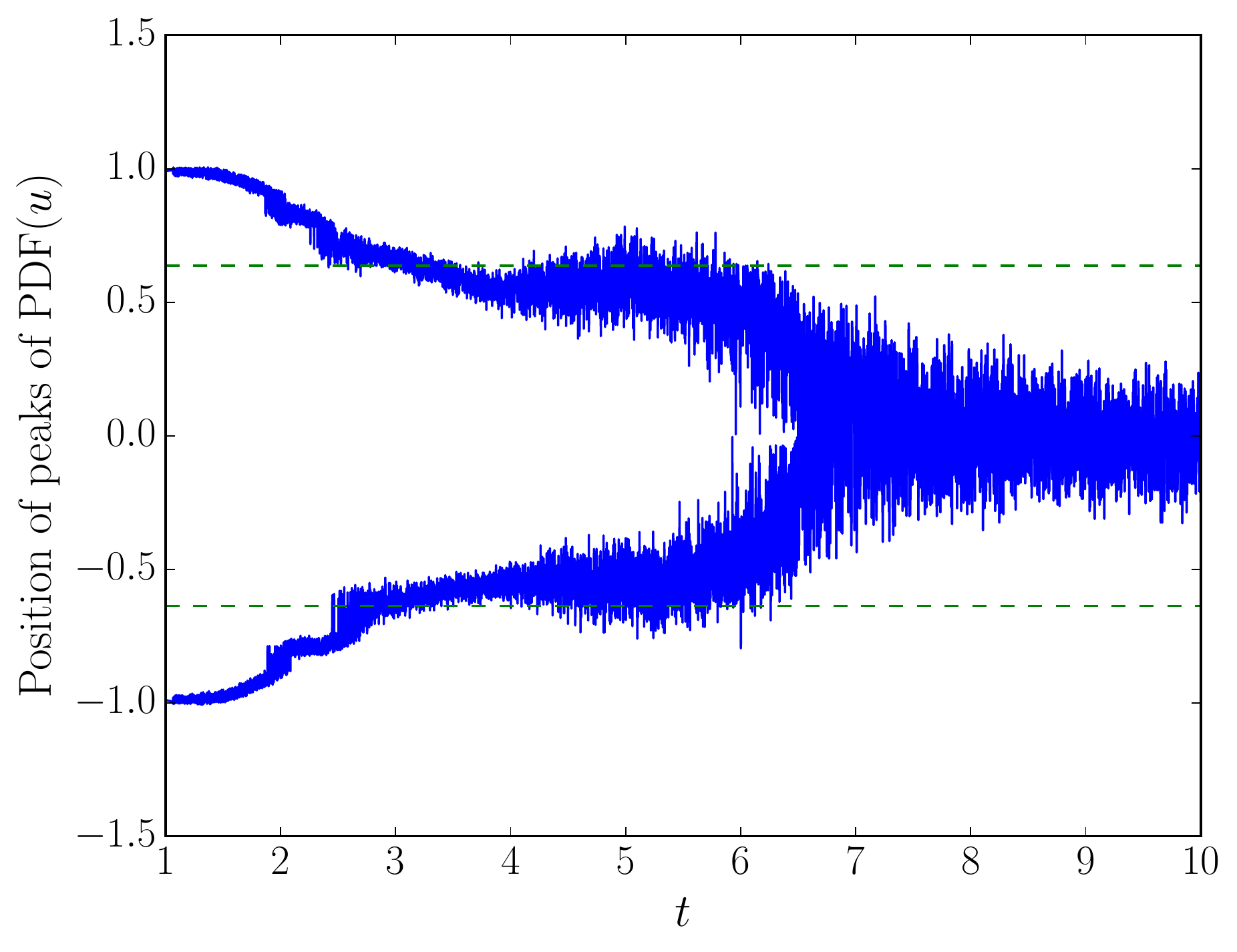}
    \caption{({\it Color online}) Evolution of the the position of the
      peaks of the histograms shown in Fig.~\ref{histevol}, as a
      function of time. It is clear how these peaks start being wide
      apart and centered around 1 and -1. Then, while the behavior of
      the PDF is that of a bimodal Gaussian distribution, these values
      fluctuate around $2/\pi$ and $-2/\pi$ (these values are
      indicated by the horizontal dashed green lines), and finally
      they vanish as equilibrium is reached in the whole domain.}
    \label{muevol}
\end{figure}

To better understand this evolution we present the spatio-temporal
plot of $u(x,t)$, and the spatio-temporal energy spectrum in
Figures~\ref{sptmp}(a) and \ref{sptmp}(b) respectively. The
spatio-temporal evolution of $u(x,t)$ is similar to those used in the
methods of characteristics, to study the formation and evolution of
shocks. Indeed, the formation of the shock at $t=1$ and at $x=\pi/2$
is clearly visible as the formation of a sharp horizontal line (note
that the shock does not propagate as it has speed $u=0$). Right after
the shock forms, a noisier horizontal line appears at $x=3\pi/2$, with
a cone (indicated by two solid lines) that widens linearly with
time. This is the tyger and the thermalized region that propagates
until covering the entire domain. Remarkably, the propagation of its
fronts has clear mean velocities  $U=\pm 2/\pi$, which is also the
slope of the two straight lines in Fig.~\ref{sptmp}(a).

The spatio-temporal spectrum has been used before to identify
structures and waves in turbulent and other complex flows
\cite{Clark14,Clark16}. It is given by
\begin{equation}
E(k,\omega) = |\hat{u} (k,w)|^2/2,
\end{equation}
where $\hat{u} (k,w)$ is the Fourier transform in time of the Fourier
coefficients of the velocity $\hat{u}(k,t)$. The Fourier transform
in time is performed from the moment the tyger appears to the time
when both partially thermalized states have developed (from $t=1$ to
$t=3$).  A flat-top window function is used to correct for the fact that
the signals are not periodic in time. Accumulation of energy near the
relation $\omega = Uk$ (with $U=2/\pi$), as observed in
Fig.~\ref{sptmp}(b), indicates that a large number of modes propagate
in real space with this velocity, confirming the observation in
Fig.~\ref{sptmp}(a).

We propose the following phenomenological argument to explain the
behavior of the tyger fronts during the transient from its formation
to the system thermalization, and to explain their observed mean
velocities. The shock and the tyger cut the total domain into two
subdomains. Thermalization is then achieved first partially in these
subdomains, and then eventually fully in the total domain. Each tyger
front is then considered as the partially thermalized solution of each
subdomain, with mean $\mu$ and variance $\sigma$ equal to the mean
velocity and mean energy inside each subdomain. So, taking the first
subdomain between $-\pi/2$ (or $3\pi/4$ in the periodic domain) and
$\pi/2$, we can get the mean velocity and mean energy directly from
the initial condition, namely
\begin{gather}
    \mu = \frac{1}{\pi} \int^{\pi/2}_{-\pi/2} \cos(x) \dif x =
    \frac{2}{\pi},
    \label{mudef1}
    \\
    \intertext{and}
    \sigma^2 + \mu^2 = \frac{1}{\pi}\int^{\pi/2}_{-\pi/2} \cos^2(x) \dif
    x = \frac{1}{2}.
    \label{sigmadef1}
\end{gather}
Note $\mu$ coincides with the observed velocity at which the fronts of
the tygers propagate.

\begin{figure}
    \centering
    \includegraphics[width=8.1cm]{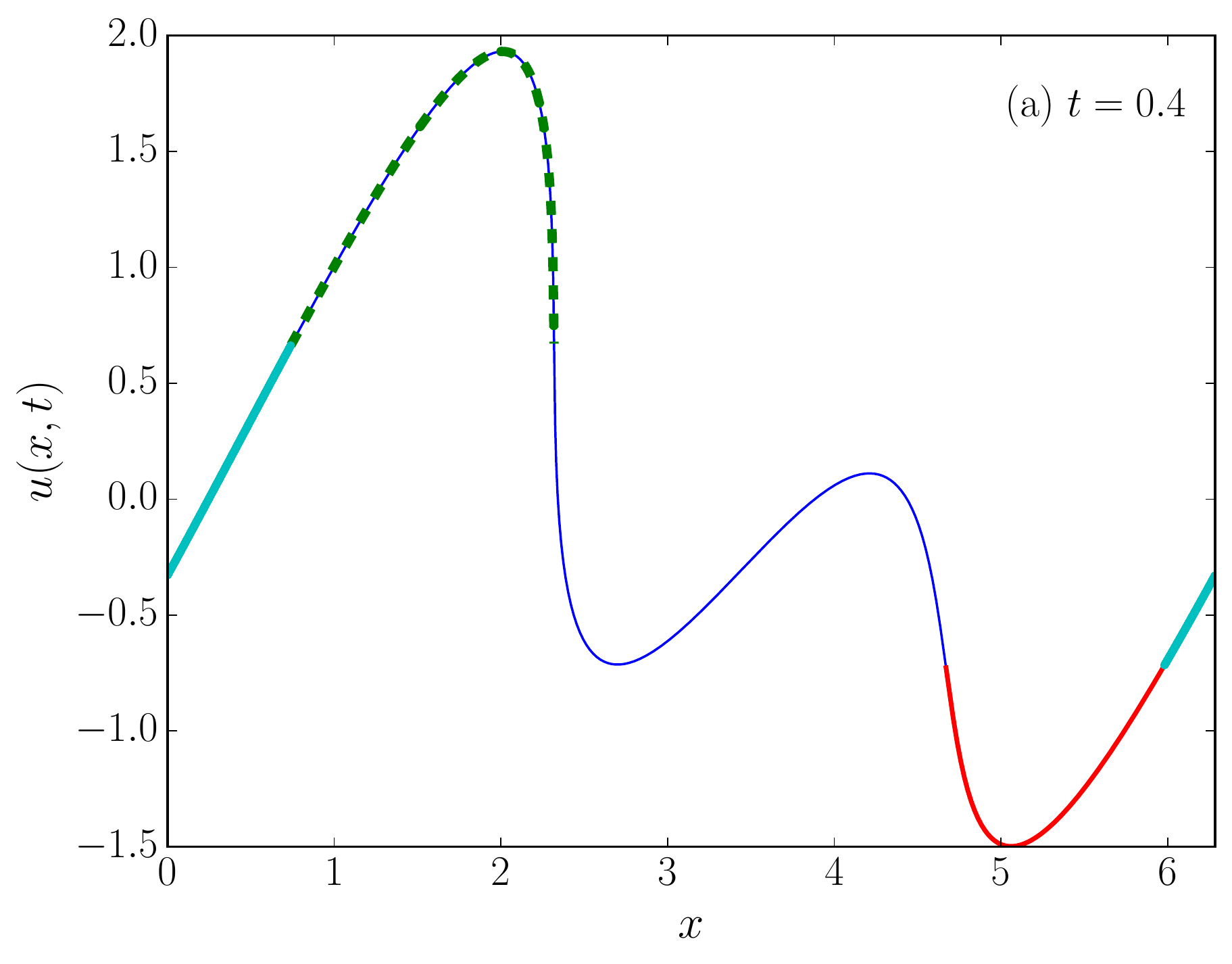}
    \includegraphics[width=8.1cm]{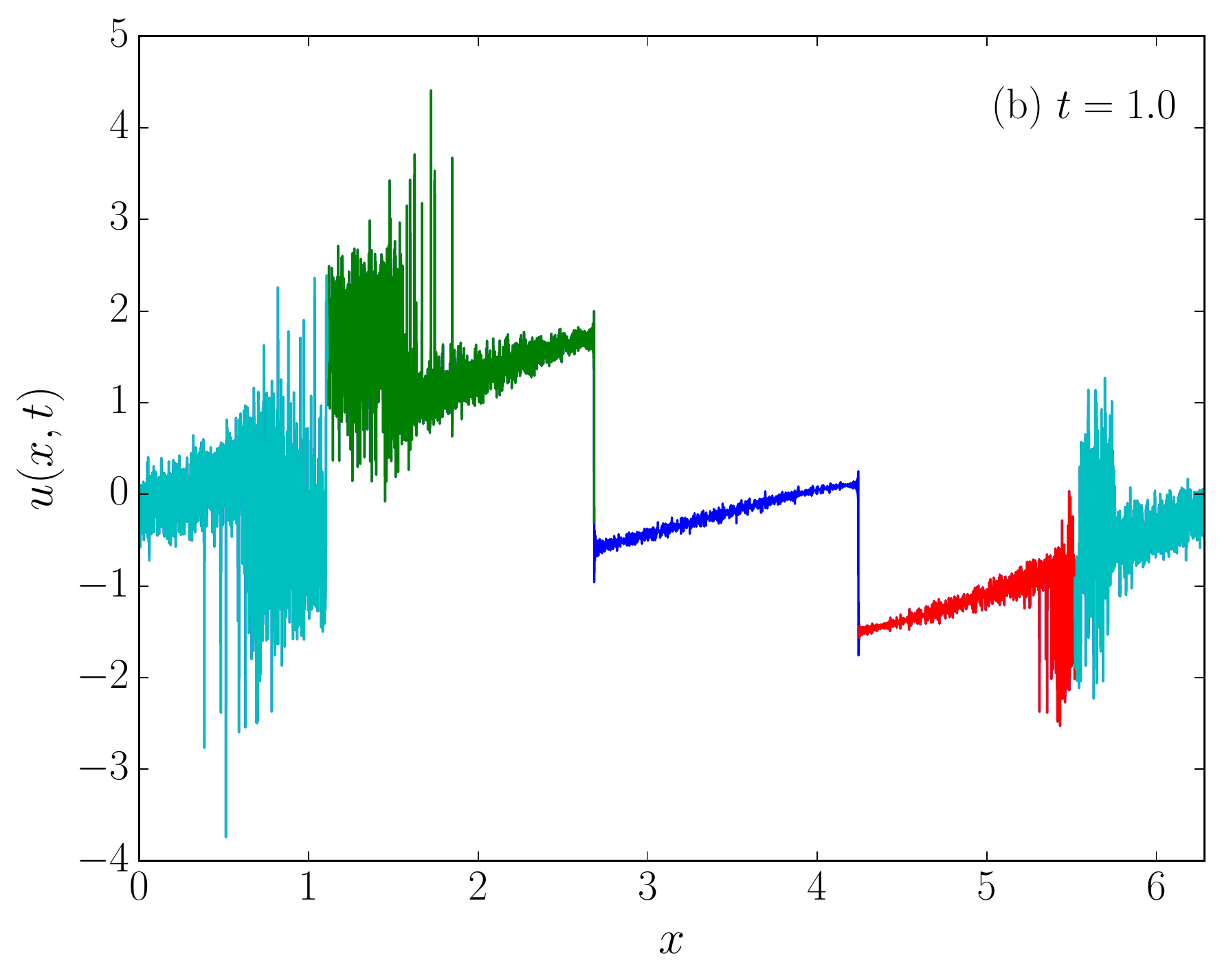}
    \includegraphics[width=8.1cm]{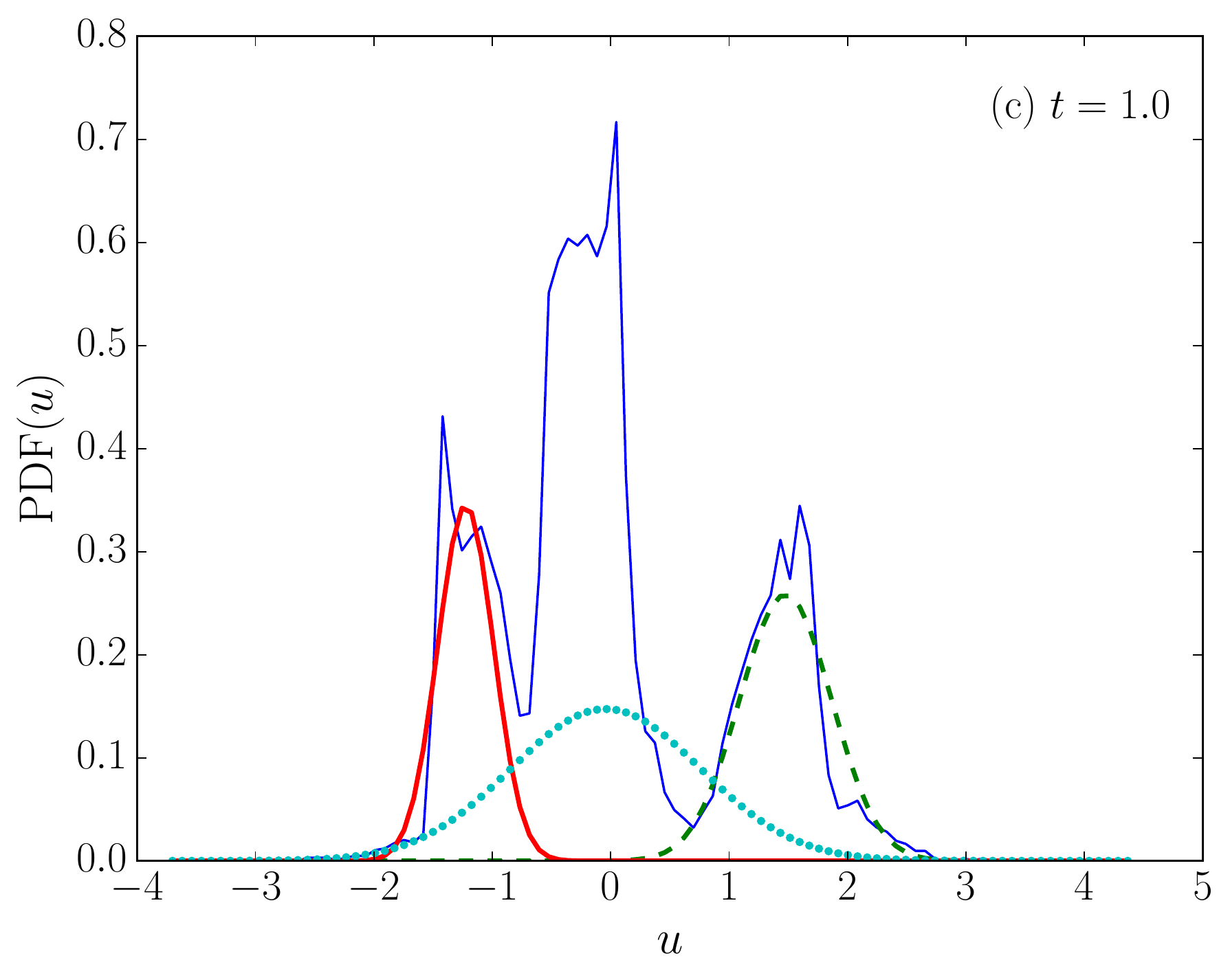}
    \caption{({\it Color online}) (a) and (b): Evolution of $u(x,t)$
      at different times in a simulation with a two-mode initial
      condition. Two shocks are formed now, and thus two tygers. 
      (c): PDF of $u(x,t=1)$ and proposed PDF for each tyger
      front. The (green) dashed line in (a) marks the region (before
      and after the appearance of the shock) contributing to the
      right peak of the PDF in (c). The mean and standard deviation of
      this region is the same as the normal distribution plotted with
      (green) dashed lines in (c). The same applies to other (colored)
      shaded regions in all three figures.}
    \label{tempevol2}
\end{figure}

\begin{figure}
    \centering
    \includegraphics[trim={5cm .5cm .8cm 0},clip,width=8.5cm]{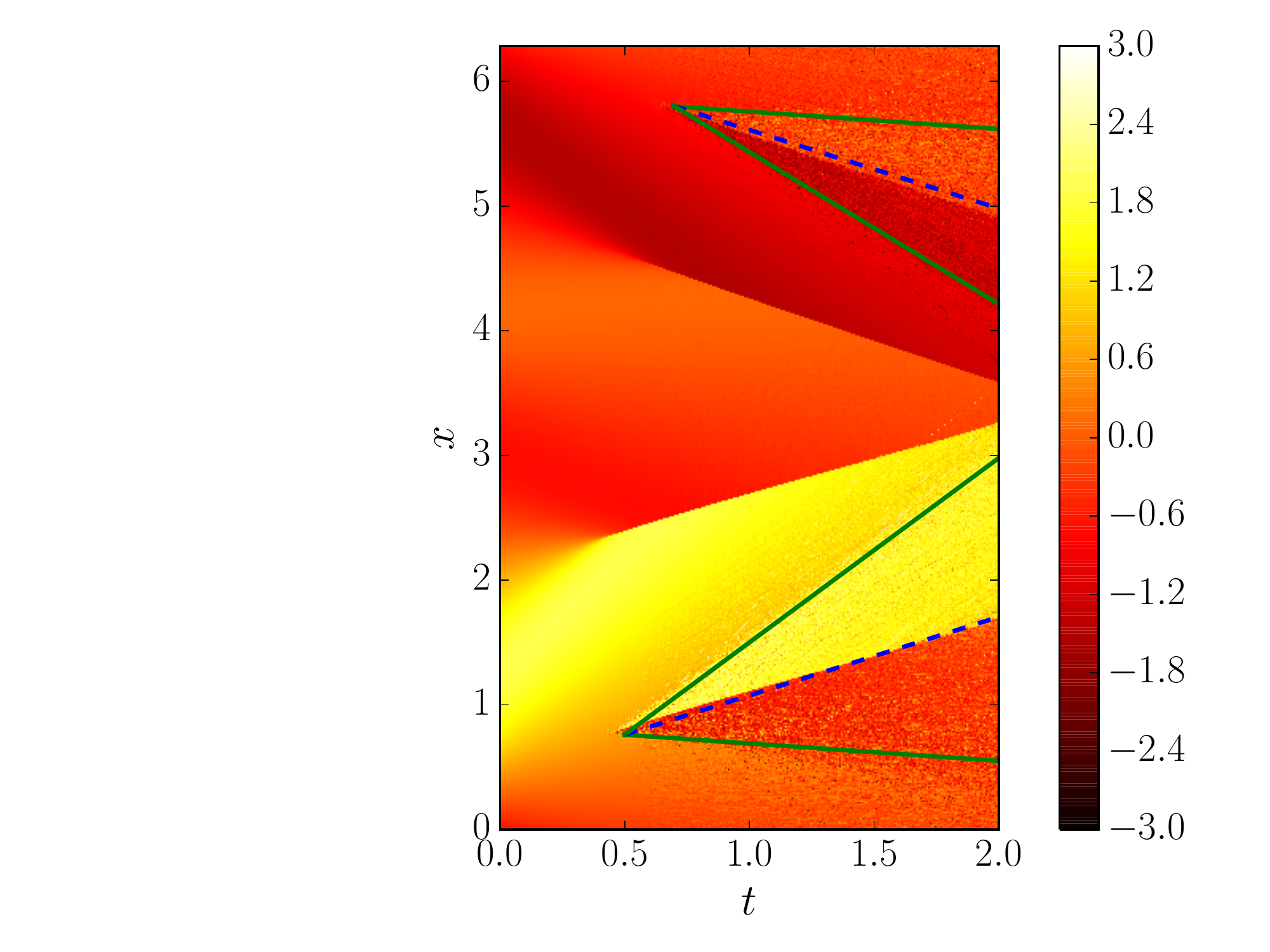}
    \includegraphics[width=8.5cm]{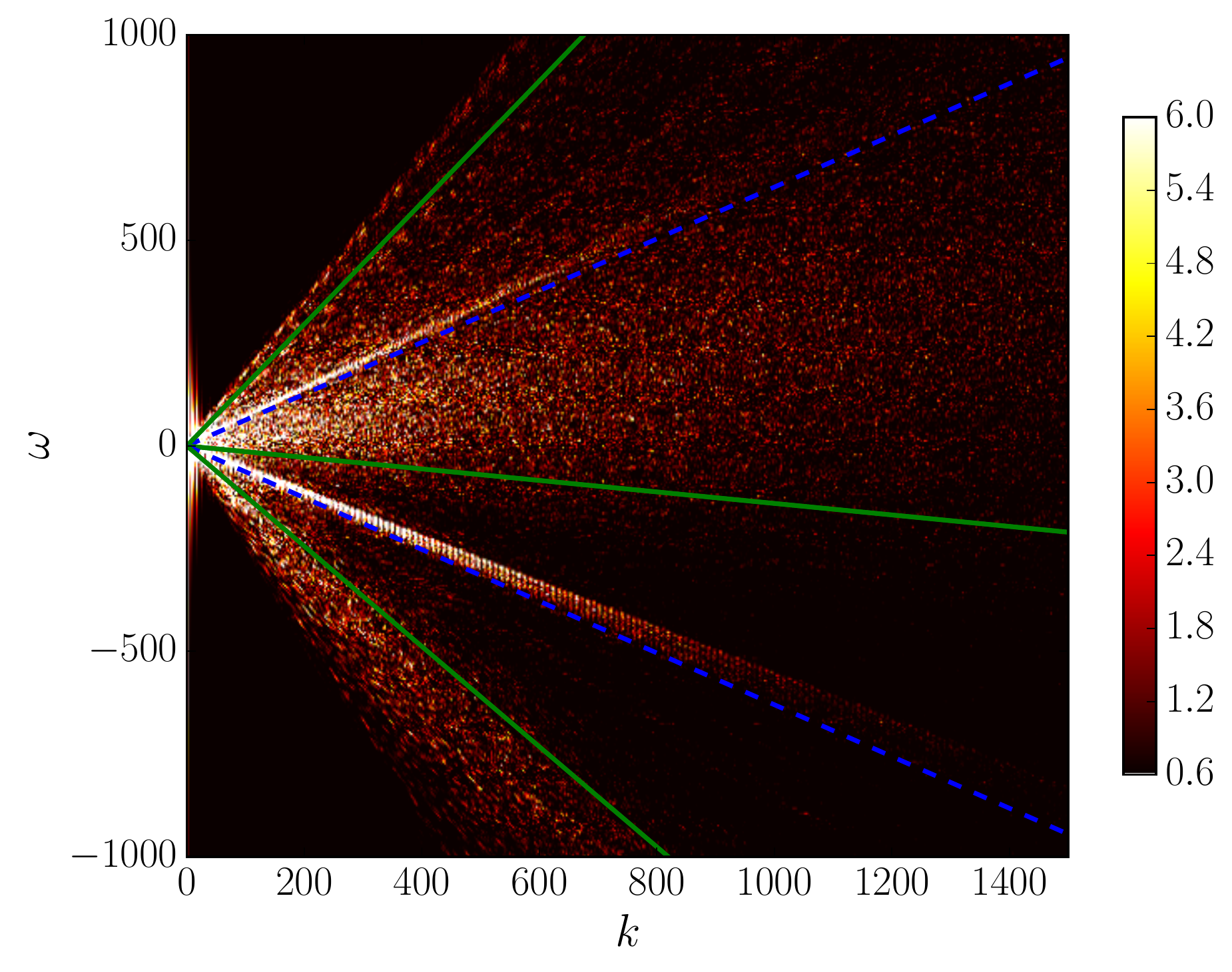}
    \caption{({\it Color online}) Spatio-temporal evolution of $u(x,t)$
    from a simulation with a two-mode initial condition: (a) Evolution
    in real space as a function of space and time, and (b) evolution in
    Fourier space as a function of frequency and wave number. The dashed
    (blue) lines correspond to the shock velocities, while the solid (green)
    lines correspond to the tygers. As two shocks form in this case, tygers
    appear from two different sites. The simple phenomenological
    theory we present is able to reproduce the mean velocity of each
    tyger.}
    \label{sptmp2}
\end{figure}

To test the validity of the assumption that each region is a partially
thermalized solution with mean and variance set by the conservation of
momentum and of energy, we compute the probability density functions 
(PDFs) of $u$ at different times; these are shown in 
Fig.~\eqref{histevol}. For early times the system has the PDF of the
cosine function, with two sharp peaks at $u=\pm 1$ as expected. At the
time of the formation of the tyger ($t=1.1$) this PDF is slightly
modified; the generation of the tyger and evolution at early times was
studied in detail in \cite{Ray11,Venkataraman17,Feng17}. For late times
($t=10$), when the system is fully thermalized, the PDF matches that
of a Gaussian with zero mean and a standard deviation of $1/2$, also
as expected from the Gibbs ensemble \cite{Majda00}. But for times in
between, the PDFs have two well defined peaks. To test that each tyger
front is close to a partially thermalized solution with mean and
standard deviation as calculated above, we plot at $t=3$ and $4.1$ in
Figs.~\ref{histevol}(c) and \ref{histevol}(d) a bimodal distribution
of the form
\begin{equation}
    \frac{1}{2 \sqrt{2\pi \sigma^2}} \left( e^{- \frac{(u-\mu)^2}{2
                \sigma^2}} + e^{- \frac{(u+\mu)^2}{2 \sigma^2}}
            \right) ,
\end{equation}
where the values of $\mu$ and $\sigma$ come from
Eqs.~\eqref{mudef1}. The proposed distributions are in good agreement
with the data, without any free parameters to improve the adjustment.
Of course, as the computation of the PDF in the simulation is done
using the entire raw data and thus mixes values from the tyger fronts
and from the non-thermalized parts of the solution, the match is not
perfect. But in spite of this, the mean and width of the peaks are
very well captured with the simple phenomenological model. Moreover,
the position and width of the peaks do not change significantly during
the transient, getting closer in time to the bimodal Gaussian
distribution (see $t=4.1$). In the Appendix, we show the
PDF of $u$ at $t=3$ using different resolutions. As stated above, our
results hold for all the resolutions studied. As the fronts of the
tygers reach $\pi/2$ from the left and the right of the shock, and
cover the entire domain (see $t=7$ in Figs.~\ref{tempevol} and
\ref{histevol}) the two peaks suddenly merge into a PDF close to
Gaussian that converges to the equilibrium solution. In order to
have a better understanding of how these distributions behave, we show
in Fig.~\ref{muevol} the time evolution of the peaks of the PDFs shown
in Fig.~\ref{histevol}. These peaks are first positioned around 1 and
$-1$, but when the partial thermalization becomes prominent we see the
values of the peaks to be around $2/\pi$ and $-2/\pi$. Eventually, the
two tyger fronts merge and the final peak is indeed centered at zero.

As an independent test we now analyze the evolution of the system
using the two-mode initial condition $u_2 (x)$ given by
Eq.~\eqref{twomode}. A snapshot of $u(x,t)$ just prior to the
formation of the shocks is shown in Fig.~\ref{tempevol2}(a), and
another one after two tygers have formed is shown in
Fig.~\ref{tempevol2}(b). As this initial condition generates two 
shocks, one tyger is formed for each shock. The points where the
shocks form are those that have $\partial_x u<0$ and 
$\partial^2_{xx} u = 0$. The tygers form far away from the shocks, at
the points of the solution with positive strain that move with the
same velocity as each shock (note the shocks in this case move with 
$u \neq 0$). The velocity of each shock can be obtained just by
inspection of the value of $u$ at the points where the shocks are, and
the point with the same velocity but with $\partial_x u>0$ in
Fig.~\ref{tempevol2}(a) is indeed the point where each tyger appears;
see also Fig.~\ref{tempevol2}(b) to see the tygers after the collapse. 

Following the previous argument, the tygers and the shocks separate
the flow in four regions (marked in different line styles and colors
in Fig.~\ref{tempevol2}): one region to the left of each tyger until
the nearest shock or tyger takes place, another region to the right of 
each tyger until the nearest shock or tyger takes place, and a region
in the center of the domain bounded by two shocks. This region shows
less noise, while all the other regions show signs of partial
thermalization as the fronts of the tygers propagate. Figure
\ref{tempevol2}(c) shows the PDF of the velocity field at $t=1$. In
the PDF of the data, three peaks are present. Superimposed to this PDF
are shown three PDFs obtained using the same methodology as described
before. The proposed PDF on the left is a Gaussian distribution with
$\mu$ and $\sigma$ given by the mean velocity and energy of the
initial conditions in the region marked by the thick (red) line (this
region has mean velocity $\mu=-1.22$ and a standard deviation of
$0.24$). The proposed PDF on the right is a Gaussian distribution with
$\mu$ and $\sigma$ obtained from the initial condition in the region
marked with the (green) dashed line (with mean velocity $1.48$ and
standard deviation of $0.39$). And the proposed PDF on the middle
corresponds to the region between the two tygers, wich has mean  
velocity equal to zero; the values of $\mu$ and $\sigma$ of the 
Gaussian distribution were obtained from the initial conditions in the
region indicated by the (cyan) thickest line. Amplitudes of the
Gaussian distributions are proportional to the area covered by each
region. The two Gaussian distributions on the left and right are in
good agreement with the data; the difference between the big center
peak in the middle and the proposed PDF is the contribution of the
region between the two fronts which is isolated from the tygers.

As an independent test and as a way to better understand the
dynamics of the solution with the two-mode initial condition, we show
the spatio-temporal plot of $u(x,t)$ and the corresponding
spatio-temporal spectrum in Fig.~\ref{sptmp2}. The dashed blue lines
correspond to the velocity of the shocks, while the solid green lines
are the mean velocities of the tygers. The regions marked in cyan,
green and red in Fig.~\ref{tempevol2}(b), where the average is taken
to calculate the mean and variance of each tyger, are the regions
between the dashed and the solid lines in Fig.~\ref{sptmp2}(a). The
evolution of the shocks and the tygers stemming from them (as well as
their mean velocities, indicated by the slopes of the straight lines)
can be clearly seen.

Similar results were obtained in tests using different initial
conditions that give rise to three or more shocks. The results thus
confirm that each region bounded by shocks or by tygers goes through a
partial thermalization, and that its properties (as well as the mean
velocity at which the front of the tyger propagates) can be obtained
from the available momentum and energy in the same region at $t=0$.
The fronts (and the tygers) act separating regions with different
thermodynamical properties (i.e., with different values of $\mu$ and
$\sigma$). As the shocks and the tygers propagate through the entire
domain, and as the fronts meet, the system finally reaches the
thermalized equilibrium described by a unique Gaussian distribution
function for the velocity, with a value of $\mu$ and $\sigma$ for the
entire domain.

\section{Conclusions} 

The transition from deterministic solutions to stochastic thermalized 
equilibria in spectrally truncated hydrodynamics problems is still
riddled with open questions. Its study can give new insights in
problems such as the development of singularities in inviscid flows,
or the development of new numerical methods for the integration of
ideal equations if the growth of thermalized solutions can be
controlled or removed. For the case of the Burgers equation, a simple
model which still exhibits remarkably complex behavior and which is
often used as a toy model of turbulence, its long-time solutions are
known \cite{Majda00}, and the triggering of the thermalization
has been understood through the discovery of the so-called tygers
\cite{Ray11}.

In this work we considered the intermediate time evolution of the
tygers, after their formation, and before the system reaches the
thermalized regime. While in previous works (see, e.g.,
\cite{Cichowlas05}) it was found that thermalization takes place
gradually in Fourier space, here we found for the Burgers equation
that in real space the two phases (thermalized and a non-thermalized)
coexist with well defined regions separated by shocks and tygers. The
propagation of the tygers, which take place with a well defined mean
velocity, results in the growth of the partially thermalized regions
until the system reaches the equilibrium solution. Moreover, the mean
velocity of propagation of the fronts, as well as the thermal
properties of each subdomain, can be obtained from the conservation of
the momentum and of the energy in each region.

\appendix*
\section{Comparison of simulations with different resolutions} 

\begin{figure*}
    \centering
    \includegraphics[width=0.3\textwidth]{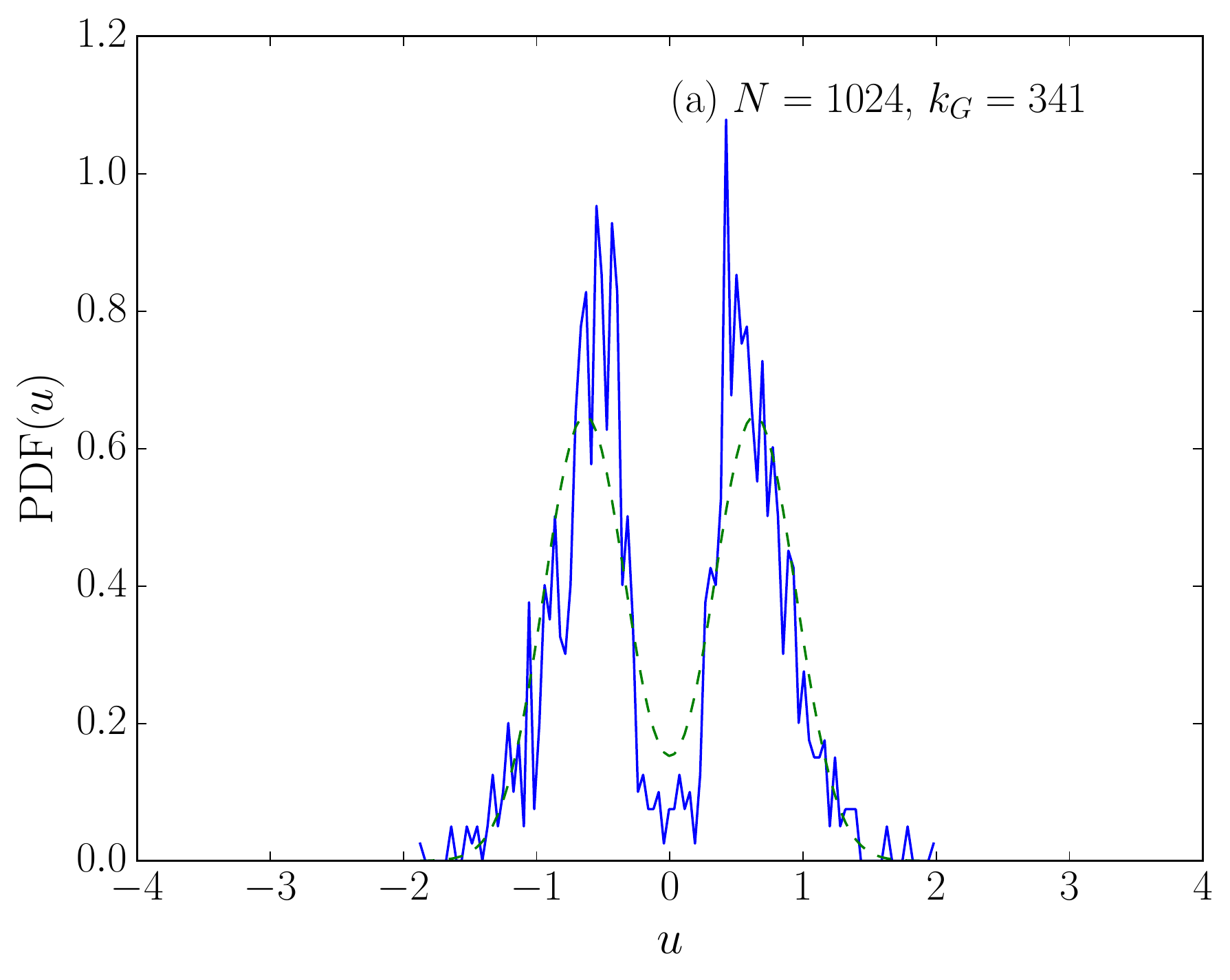}
    \includegraphics[width=0.3\textwidth]{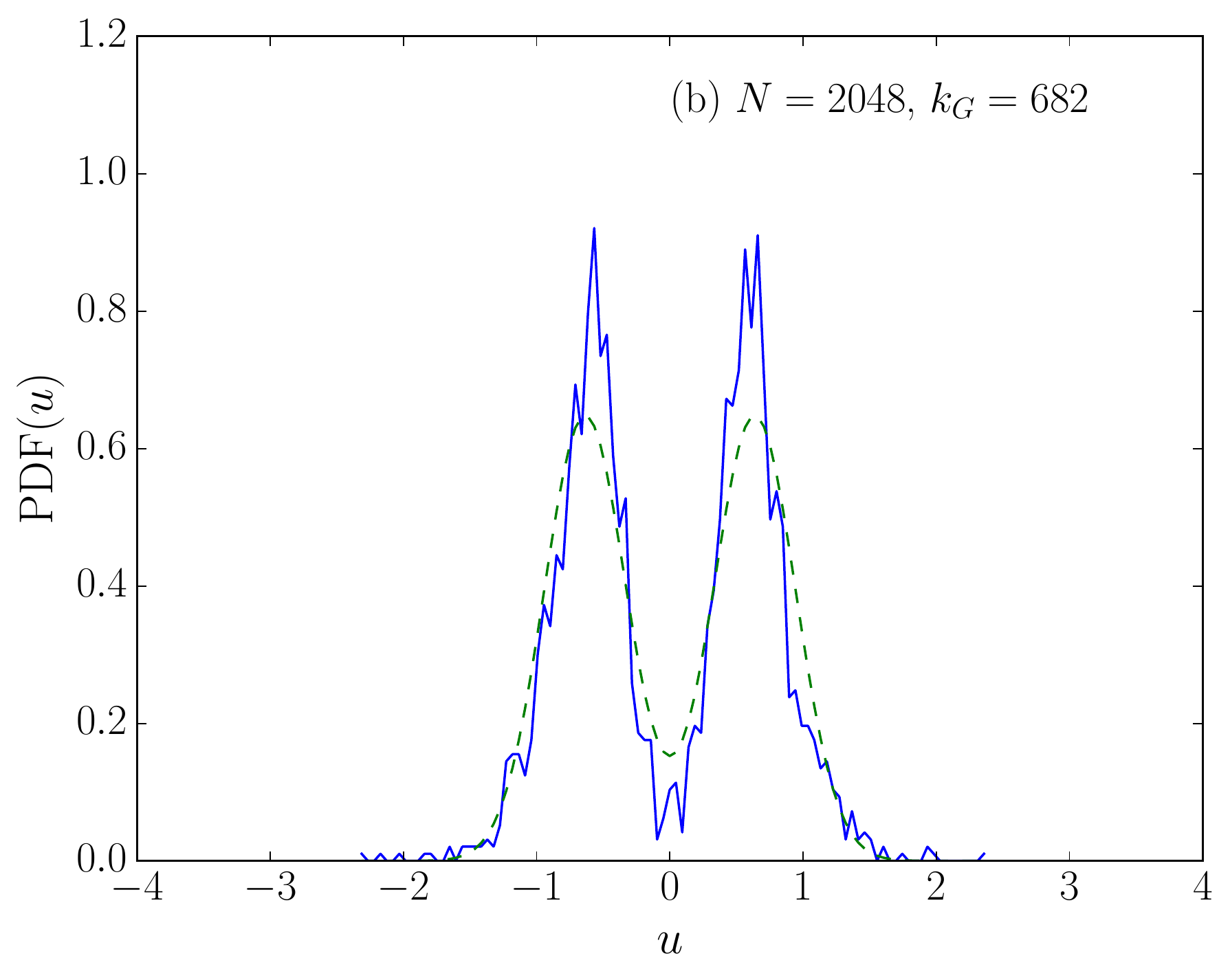}
    \includegraphics[width=0.3\textwidth]{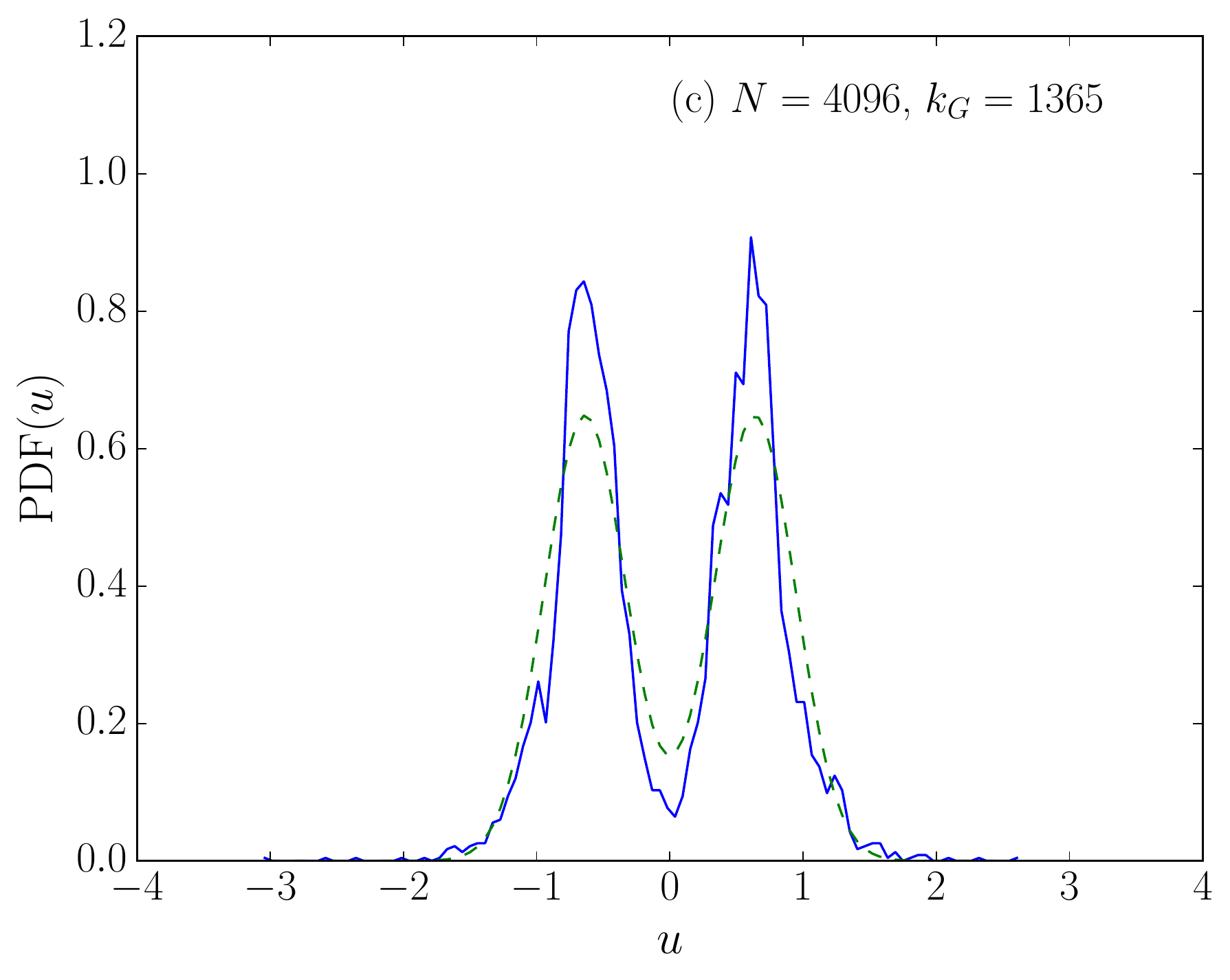}
    \includegraphics[width=0.3\textwidth]{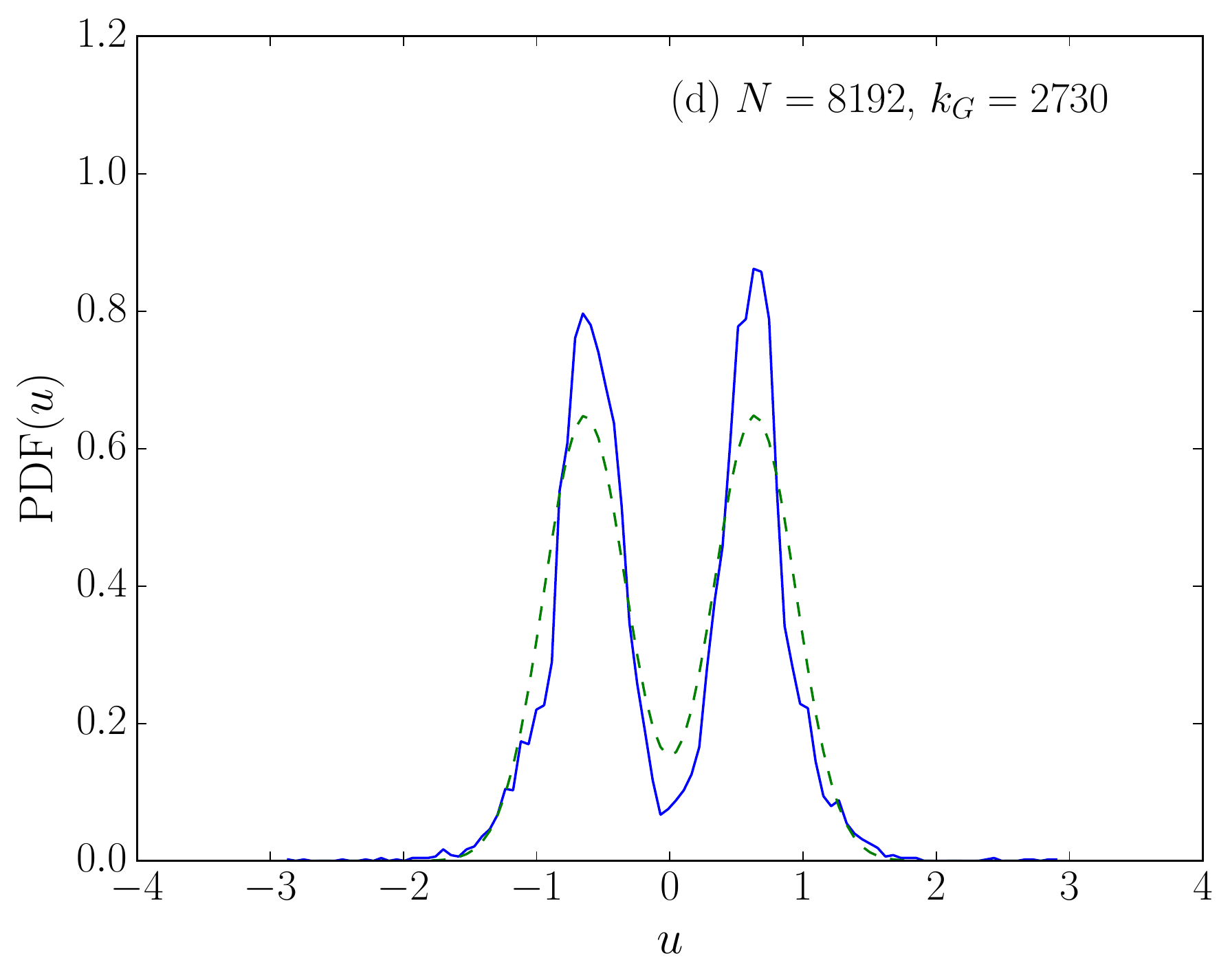}
    \includegraphics[width=0.3\textwidth]{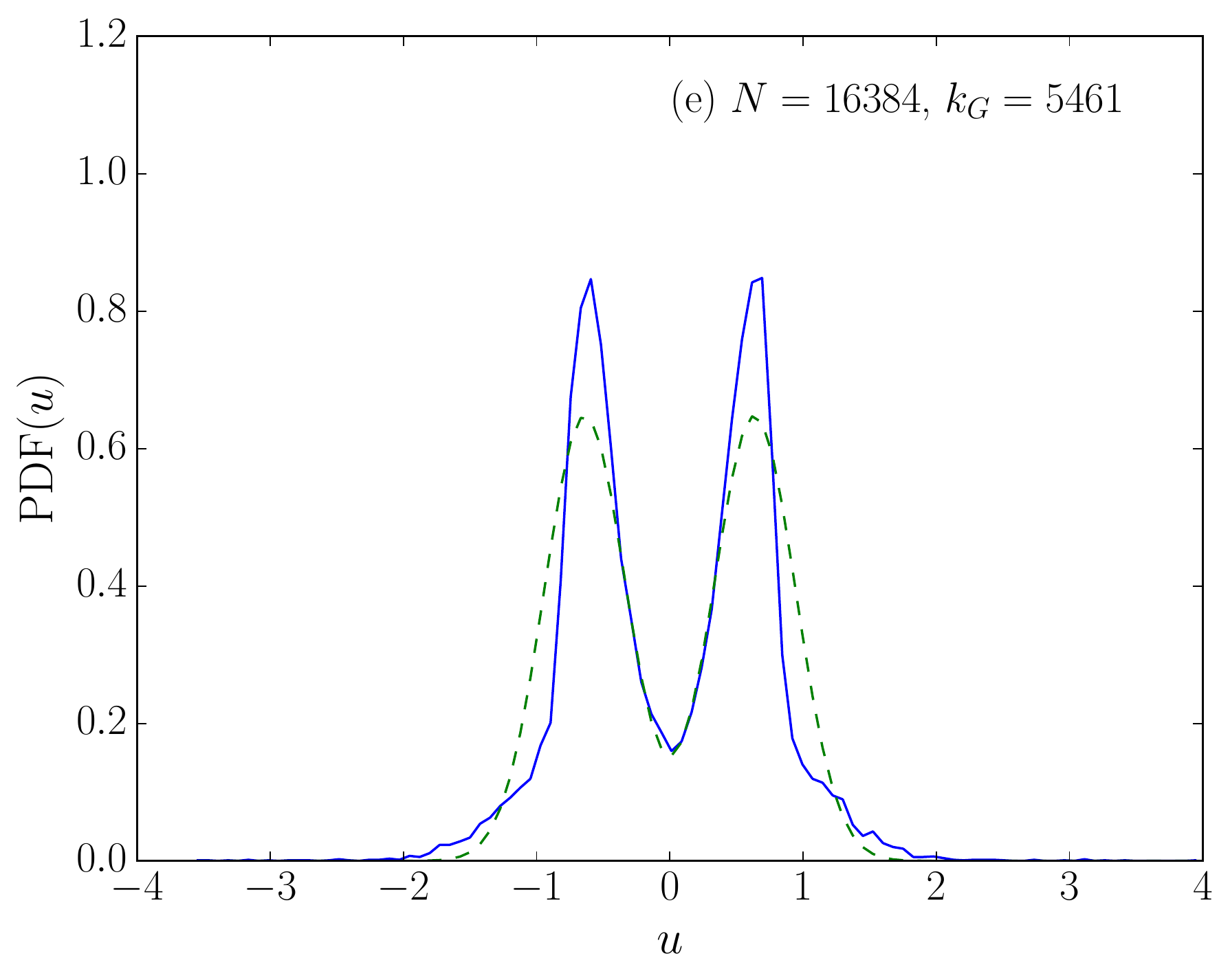}
    \includegraphics[width=0.3\textwidth]{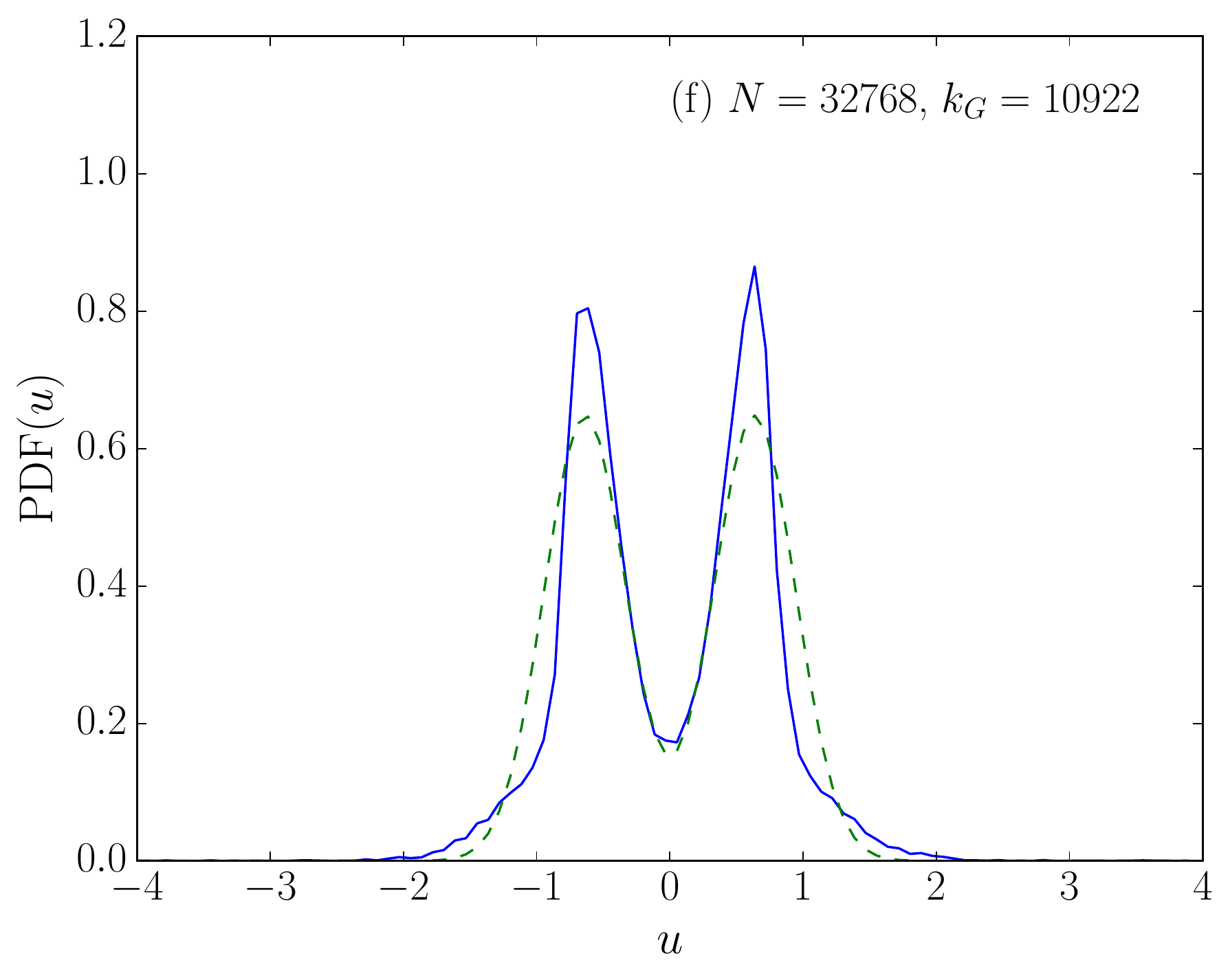}
    \caption{({\it Color online}) Probability density function (PDF) of
    $u(x,t)$ at $t=3$ for different simulation in (blue) solid lines; in
    (green) dashed lines is the proposed PDF. All simulations have the
    same initial condition but different resolutuions. The bimodal
    Gaussian distribution plotted is the same for every case. Our
    results are independent of resolution for the resolutions studied.}
    \label{resolutions}
\end{figure*}

In Fig.~\ref{resolutions} we show the PDF of $u$ at $t=3$ for different
simulations, all with the same initial condition but with different
spatial resolution. In all cases we plot the same bimodal Gaussian
distribution (with the same parameters $\mu$ and $\sigma$ coming from
Eqs.~\eqref{mudef1} and \eqref{sigmadef1}) as in Figs.~\ref{histevol}(c)
and (d), and in all cases it properly describes the behavior of the PDF.
Aside from the more ragged looks of the lower resolution simulations,
which is just the result of having less data points to construct the
histograms, all simulations display the same dynamics.

\begin{acknowledgments}
The authors acknowledge financial support from Grant No.~ECOS-Sud
A13E01. P.C.dL. and P.D.M. acknowledge support from UBACYT Grant
No.~20020130100738BA, and PICT  Grants Nos.~2011-1529 and 
2015-3530. P.C.dL. acknowledges funding from the European Research
Council under the European Community's Seventh Framework Program, ERC
Grant Agreement No.~339032.
\end{acknowledgments} 

\bibliography{ms}

\begin{thebibliography}{27}%
\makeatletter
\providecommand \@ifxundefined [1]{%
 \@ifx{#1\undefined}
}%
\providecommand \@ifnum [1]{%
 \ifnum #1\expandafter \@firstoftwo
 \else \expandafter \@secondoftwo
 \fi
}%
\providecommand \@ifx [1]{%
 \ifx #1\expandafter \@firstoftwo
 \else \expandafter \@secondoftwo
 \fi
}%
\providecommand \natexlab [1]{#1}%
\providecommand \enquote  [1]{``#1''}%
\providecommand \bibnamefont  [1]{#1}%
\providecommand \bibfnamefont [1]{#1}%
\providecommand \citenamefont [1]{#1}%
\providecommand \href@noop [0]{\@secondoftwo}%
\providecommand \href [0]{\begingroup \@sanitize@url \@href}%
\providecommand \@href[1]{\@@startlink{#1}\@@href}%
\providecommand \@@href[1]{\endgroup#1\@@endlink}%
\providecommand \@sanitize@url [0]{\catcode `\\12\catcode `\$12\catcode
  `\&12\catcode `\#12\catcode `\^12\catcode `\_12\catcode `\%12\relax}%
\providecommand \@@startlink[1]{}%
\providecommand \@@endlink[0]{}%
\providecommand \url  [0]{\begingroup\@sanitize@url \@url }%
\providecommand \@url [1]{\endgroup\@href {#1}{\urlprefix }}%
\providecommand \urlprefix  [0]{URL }%
\providecommand \Eprint [0]{\href }%
\providecommand \doibase [0]{http://dx.doi.org/}%
\providecommand \selectlanguage [0]{\@gobble}%
\providecommand \bibinfo  [0]{\@secondoftwo}%
\providecommand \bibfield  [0]{\@secondoftwo}%
\providecommand \translation [1]{[#1]}%
\providecommand \BibitemOpen [0]{}%
\providecommand \bibitemStop [0]{}%
\providecommand \bibitemNoStop [0]{.\EOS\space}%
\providecommand \EOS [0]{\spacefactor3000\relax}%
\providecommand \BibitemShut  [1]{\csname bibitem#1\endcsname}%
\let\auto@bib@innerbib\@empty
\bibitem [{\citenamefont {Kraichnan}\ and\ \citenamefont
  {Chen}(1989)}]{Kraichnan89}%
  \BibitemOpen
  \bibfield  {author} {\bibinfo {author} {\bibfnamefont {Robert~H.}\
  \bibnamefont {Kraichnan}}\ and\ \bibinfo {author} {\bibfnamefont {Shiyi}\
  \bibnamefont {Chen}},\ }\bibfield  {title} {\enquote {\bibinfo {title} {Is
  there a statistical mechanics of turbulence?}}\ }\href {\doibase
  10.1016/0167-2789(89)90126-7} {\bibfield  {journal} {\bibinfo  {journal}
  {Physica D: Nonlinear Phenomena}\ }\textbf {\bibinfo {volume} {37}},\
  \bibinfo {pages} {160--172} (\bibinfo {year} {1989})}\BibitemShut {NoStop}%
\bibitem [{\citenamefont {Lee}(1952)}]{Lee52}%
  \BibitemOpen
  \bibfield  {author} {\bibinfo {author} {\bibfnamefont {T.~D.}\ \bibnamefont
  {Lee}},\ }\bibfield  {title} {\enquote {\bibinfo {title} {{On} {some}
  {statistical} {properties} {of} {hydrodynamical} {and}
  {magneto}-{hydrodynamical} {fields}},}\ }\href@noop {} {\bibfield  {journal}
  {\bibinfo  {journal} {Quarterly of Applied Mathematics}\ }\textbf {\bibinfo
  {volume} {10}},\ \bibinfo {pages} {69--74} (\bibinfo {year}
  {1952})}\BibitemShut {NoStop}%
\bibitem [{\citenamefont {Kraichnan}(1967)}]{Kraichnan67}%
  \BibitemOpen
  \bibfield  {author} {\bibinfo {author} {\bibfnamefont {Robert~H.}\
  \bibnamefont {Kraichnan}},\ }\bibfield  {title} {\enquote {\bibinfo {title}
  {Inertial {Ranges} in {Two}-{Dimensional} {Turbulence}},}\ }\href {\doibase
  10.1063/1.1762301} {\bibfield  {journal} {\bibinfo  {journal} {Physics of
  Fluids}\ }\textbf {\bibinfo {volume} {10}},\ \bibinfo {pages} {1417--1423}
  (\bibinfo {year} {1967})}\BibitemShut {NoStop}%
\bibitem [{\citenamefont {Ting}\ \emph {et~al.}(1986)\citenamefont {Ting},
  \citenamefont {Matthaeus},\ and\ \citenamefont {Montgomery}}]{Ting86}%
  \BibitemOpen
  \bibfield  {author} {\bibinfo {author} {\bibfnamefont {A.~C.}\ \bibnamefont
  {Ting}}, \bibinfo {author} {\bibfnamefont {W.~H.}\ \bibnamefont {Matthaeus}},
  \ and\ \bibinfo {author} {\bibfnamefont {D.}~\bibnamefont {Montgomery}},\
  }\bibfield  {title} {\enquote {\bibinfo {title} {Turbulent relaxation
  processes in magnetohydrodynamics},}\ }\href@noop {} {\bibfield  {journal}
  {\bibinfo  {journal} {Phys.\ Fluids}\ }\textbf {\bibinfo {volume} {29}},\
  \bibinfo {pages} {3261} (\bibinfo {year} {1986})}\BibitemShut {NoStop}%
\bibitem [{\citenamefont {Cichowlas}\ \emph {et~al.}(2005)\citenamefont
  {Cichowlas}, \citenamefont {Bonaïti}, \citenamefont {Debbasch},\ and\
  \citenamefont {Brachet}}]{Cichowlas05}%
  \BibitemOpen
  \bibfield  {author} {\bibinfo {author} {\bibfnamefont {Cyril}\ \bibnamefont
  {Cichowlas}}, \bibinfo {author} {\bibfnamefont {Pauline}\ \bibnamefont
  {Bonaïti}}, \bibinfo {author} {\bibfnamefont {Fabrice}\ \bibnamefont
  {Debbasch}}, \ and\ \bibinfo {author} {\bibfnamefont {Marc}\ \bibnamefont
  {Brachet}},\ }\bibfield  {title} {\enquote {\bibinfo {title} {Effective
  {Dissipation} and {Turbulence} in {Spectrally} {Truncated} {Euler}
  {Flows}},}\ }\href {\doibase 10.1103/PhysRevLett.95.264502} {\bibfield
  {journal} {\bibinfo  {journal} {Phys. Rev. Lett.}\ }\textbf {\bibinfo
  {volume} {95}},\ \bibinfo {pages} {264502} (\bibinfo {year}
  {2005})}\BibitemShut {NoStop}%
\bibitem [{\citenamefont {Krstulovic}\ \emph
  {et~al.}(2009{\natexlab{a}})\citenamefont {Krstulovic}, \citenamefont
  {Mininni}, \citenamefont {Brachet},\ and\ \citenamefont
  {Pouquet}}]{Krstulovic09b}%
  \BibitemOpen
  \bibfield  {author} {\bibinfo {author} {\bibfnamefont {G.}~\bibnamefont
  {Krstulovic}}, \bibinfo {author} {\bibfnamefont {P.~D.}\ \bibnamefont
  {Mininni}}, \bibinfo {author} {\bibfnamefont {M.~E.}\ \bibnamefont
  {Brachet}}, \ and\ \bibinfo {author} {\bibfnamefont {A.}~\bibnamefont
  {Pouquet}},\ }\bibfield  {title} {\enquote {\bibinfo {title} {Cascades,
  thermalization, and eddy viscosity in helical {Galerkin} truncated {Euler}
  flows},}\ }\href {\doibase 10.1103/PhysRevE.79.056304} {\bibfield  {journal}
  {\bibinfo  {journal} {Phys.\ Rev.\ E}\ }\textbf {\bibinfo {volume} {79}},\
  \bibinfo {pages} {056304} (\bibinfo {year} {2009}{\natexlab{a}})}\BibitemShut
  {NoStop}%
\bibitem [{\citenamefont {Krstulovic}\ \emph {et~al.}(2011)\citenamefont
  {Krstulovic}, \citenamefont {Brachet},\ and\ \citenamefont
  {Pouquet}}]{Krstulovic11}%
  \BibitemOpen
  \bibfield  {author} {\bibinfo {author} {\bibfnamefont {Giorgio}\ \bibnamefont
  {Krstulovic}}, \bibinfo {author} {\bibfnamefont {Marc-Etienne}\ \bibnamefont
  {Brachet}}, \ and\ \bibinfo {author} {\bibfnamefont {Annick}\ \bibnamefont
  {Pouquet}},\ }\bibfield  {title} {\enquote {\bibinfo {title} {Alfv\'en waves
  and ideal two-dimensional {Galerkin} truncated magnetohydrodynamics},}\
  }\href {\doibase 10.1103/PhysRevE.84.016410} {\bibfield  {journal} {\bibinfo
  {journal} {Phys.\ Rev.\ E}\ }\textbf {\bibinfo {volume} {84}},\ \bibinfo
  {pages} {016410} (\bibinfo {year} {2011})}\BibitemShut {NoStop}%
\bibitem [{\citenamefont {Krstulovic}\ \emph
  {et~al.}(2009{\natexlab{b}})\citenamefont {Krstulovic}, \citenamefont
  {Cartes}, \citenamefont {Brachet},\ and\ \citenamefont
  {Tirapegui}}]{Krstulovic09}%
  \BibitemOpen
  \bibfield  {author} {\bibinfo {author} {\bibfnamefont {Giorgio}\ \bibnamefont
  {Krstulovic}}, \bibinfo {author} {\bibfnamefont {Carlos}\ \bibnamefont
  {Cartes}}, \bibinfo {author} {\bibfnamefont {Marc}\ \bibnamefont {Brachet}},
  \ and\ \bibinfo {author} {\bibfnamefont {Enrique}\ \bibnamefont
  {Tirapegui}},\ }\bibfield  {title} {\enquote {\bibinfo {title} {Generation
  and characterization of absolute equilibrium of compressible flows},}\ }\href
  {\doibase 10.1142/S021812740902489X} {\bibfield  {journal} {\bibinfo
  {journal} {International Journal of Bifurcation and Chaos}\ }\textbf
  {\bibinfo {volume} {19}},\ \bibinfo {pages} {3445--3459} (\bibinfo {year}
  {2009}{\natexlab{b}})}\BibitemShut {NoStop}%
\bibitem [{\citenamefont {Krstulovic}\ and\ \citenamefont
  {Brachet}(2011)}]{Krstulovic11b}%
  \BibitemOpen
  \bibfield  {author} {\bibinfo {author} {\bibfnamefont {Giorgio}\ \bibnamefont
  {Krstulovic}}\ and\ \bibinfo {author} {\bibfnamefont {Marc}\ \bibnamefont
  {Brachet}},\ }\bibfield  {title} {\enquote {\bibinfo {title} {Energy cascade
  with small-scale thermalization, counterflow metastability, and anomalous
  velocity of vortex rings in {Fourier}-truncated {Gross}-{Pitaevskii}
  equation},}\ }\href {\doibase 10.1103/PhysRevE.83.066311} {\bibfield
  {journal} {\bibinfo  {journal} {Phys.\ Rev.\ E}\ }\textbf {\bibinfo {volume}
  {83}},\ \bibinfo {pages} {066311} (\bibinfo {year} {2011})}\BibitemShut
  {NoStop}%
\bibitem [{\citenamefont {Zhu}(2016)}]{Zhu16}%
  \BibitemOpen
  \bibfield  {author} {\bibinfo {author} {\bibfnamefont {Jian-Zhou}\
  \bibnamefont {Zhu}},\ }\bibfield  {title} {\enquote {\bibinfo {title}
  {Isotropic polarization of compressible flows},}\ }\href {\doibase
  10.1017/jfm.2015.692} {\bibfield  {journal} {\bibinfo  {journal} {J.\ Fluid
  Mech.}\ }\textbf {\bibinfo {volume} {787}},\ \bibinfo {pages} {440--448}
  (\bibinfo {year} {2016})}\BibitemShut {NoStop}%
\bibitem [{\citenamefont {Zhu}\ and\ \citenamefont {Hammett}(2010)}]{Zhu10}%
  \BibitemOpen
  \bibfield  {author} {\bibinfo {author} {\bibfnamefont {Jian-Zhou}\
  \bibnamefont {Zhu}}\ and\ \bibinfo {author} {\bibfnamefont {Gregory~W.}\
  \bibnamefont {Hammett}},\ }\bibfield  {title} {\enquote {\bibinfo {title}
  {Gyrokinetic statistical absolute equilibrium and turbulence},}\ }\href
  {\doibase 10.1063/1.3514141} {\bibfield  {journal} {\bibinfo  {journal}
  {Physics of Plasmas}\ }\textbf {\bibinfo {volume} {17}},\ \bibinfo {pages}
  {122307} (\bibinfo {year} {2010})}\BibitemShut {NoStop}%
\bibitem [{\citenamefont {Dmitruk}\ \emph {et~al.}(2014)\citenamefont
  {Dmitruk}, \citenamefont {Mininni}, \citenamefont {Pouquet}, \citenamefont
  {Servidio},\ and\ \citenamefont {Matthaeus}}]{Dmitruk14}%
  \BibitemOpen
  \bibfield  {author} {\bibinfo {author} {\bibfnamefont {P.}~\bibnamefont
  {Dmitruk}}, \bibinfo {author} {\bibfnamefont {P.~D.}\ \bibnamefont
  {Mininni}}, \bibinfo {author} {\bibfnamefont {A.}~\bibnamefont {Pouquet}},
  \bibinfo {author} {\bibfnamefont {S.}~\bibnamefont {Servidio}}, \ and\
  \bibinfo {author} {\bibfnamefont {W.~H.}\ \bibnamefont {Matthaeus}},\
  }\bibfield  {title} {\enquote {\bibinfo {title} {Magnetic field reversals and
  long-time memory in conducting flows},}\ }\href@noop {} {\bibfield  {journal}
  {\bibinfo  {journal} {Phys.\ Rev.\ E}\ }\textbf {\bibinfo {volume} {90}},\
  \bibinfo {pages} {043010} (\bibinfo {year} {2014})}\BibitemShut {NoStop}%
\bibitem [{\citenamefont {Prasath}\ \emph {et~al.}(2014)\citenamefont
  {Prasath}, \citenamefont {Fauve},\ and\ \citenamefont {Brachet}}]{Prasath14}%
  \BibitemOpen
  \bibfield  {author} {\bibinfo {author} {\bibfnamefont {Srinivasa
  Gopalakrishnan~Ganga}\ \bibnamefont {Prasath}}, \bibinfo {author}
  {\bibfnamefont {Stéphan}\ \bibnamefont {Fauve}}, \ and\ \bibinfo {author}
  {\bibfnamefont {Marc}\ \bibnamefont {Brachet}},\ }\bibfield  {title}
  {\enquote {\bibinfo {title} {Dynamo action by turbulence in absolute
  equilibrium},}\ }\href {\doibase 10.1209/0295-5075/106/29002} {\bibfield
  {journal} {\bibinfo  {journal} {EPL (Europhysics Letters)}\ }\textbf
  {\bibinfo {volume} {106}},\ \bibinfo {pages} {29002} (\bibinfo {year}
  {2014})}\BibitemShut {NoStop}%
\bibitem [{\citenamefont {Teitelbaum}\ and\ \citenamefont
  {Mininni}(2012)}]{Teitelbaum12}%
  \BibitemOpen
  \bibfield  {author} {\bibinfo {author} {\bibfnamefont {T.}~\bibnamefont
  {Teitelbaum}}\ and\ \bibinfo {author} {\bibfnamefont {P.~D.}\ \bibnamefont
  {Mininni}},\ }\bibfield  {title} {\enquote {\bibinfo {title} {Thermalization
  and free decay in surface quasigeostrophic flows},}\ }\href@noop {}
  {\bibfield  {journal} {\bibinfo  {journal} {Phys.\ Rev.\ E}\ }\textbf
  {\bibinfo {volume} {86}},\ \bibinfo {pages} {016323} (\bibinfo {year}
  {2012})}\BibitemShut {NoStop}%
\bibitem [{\citenamefont {Ray}\ \emph {et~al.}(2011)\citenamefont {Ray},
  \citenamefont {Frisch}, \citenamefont {Nazarenko},\ and\ \citenamefont
  {Matsumoto}}]{Ray11}%
  \BibitemOpen
  \bibfield  {author} {\bibinfo {author} {\bibfnamefont {Samriddhi~Sankar}\
  \bibnamefont {Ray}}, \bibinfo {author} {\bibfnamefont {Uriel}\ \bibnamefont
  {Frisch}}, \bibinfo {author} {\bibfnamefont {Sergei}\ \bibnamefont
  {Nazarenko}}, \ and\ \bibinfo {author} {\bibfnamefont {Takeshi}\ \bibnamefont
  {Matsumoto}},\ }\bibfield  {title} {\enquote {\bibinfo {title} {Resonance
  phenomenon for the {Galerkin}-truncated {Burgers} and {Euler} equations},}\
  }\href {\doibase 10.1103/PhysRevE.84.016301} {\bibfield  {journal} {\bibinfo
  {journal} {Phys. Rev. E}\ }\textbf {\bibinfo {volume} {84}},\ \bibinfo
  {pages} {016301} (\bibinfo {year} {2011})}\BibitemShut {NoStop}%
\bibitem [{\citenamefont {Lax}\ and\ \citenamefont {Levermore}(1979)}]{Lax79}%
  \BibitemOpen
  \bibfield  {author} {\bibinfo {author} {\bibfnamefont {Peter~D.}\
  \bibnamefont {Lax}}\ and\ \bibinfo {author} {\bibfnamefont {C.~David}\
  \bibnamefont {Levermore}},\ }\bibfield  {title} {\enquote {\bibinfo {title}
  {The zero dispersion limit for the {Korteweg}-{deVries} {KdV} equation},}\
  }\href@noop {} {\bibfield  {journal} {\bibinfo  {journal} {PNAS}\ }\textbf
  {\bibinfo {volume} {76}},\ \bibinfo {pages} {3602--3606} (\bibinfo {year}
  {1979})}\BibitemShut {NoStop}%
\bibitem [{\citenamefont {Goodman}\ and\ \citenamefont
  {Lax}(1988)}]{Goodman88}%
  \BibitemOpen
  \bibfield  {author} {\bibinfo {author} {\bibfnamefont {Jonathan}\
  \bibnamefont {Goodman}}\ and\ \bibinfo {author} {\bibfnamefont {Peter~D.}\
  \bibnamefont {Lax}},\ }\bibfield  {title} {\enquote {\bibinfo {title} {On
  dispersive difference schemes. {I}},}\ }\href {\doibase
  10.1002/cpa.3160410506} {\bibfield  {journal} {\bibinfo  {journal} {Comm.
  Pure Appl. Math.}\ }\textbf {\bibinfo {volume} {41}},\ \bibinfo {pages}
  {591--613} (\bibinfo {year} {1988})}\BibitemShut {NoStop}%
\bibitem [{\citenamefont {Hou}\ and\ \citenamefont {Lax}(1991)}]{Hou91}%
  \BibitemOpen
  \bibfield  {author} {\bibinfo {author} {\bibfnamefont {Thomas~Y.}\
  \bibnamefont {Hou}}\ and\ \bibinfo {author} {\bibfnamefont {Peter~D.}\
  \bibnamefont {Lax}},\ }\bibfield  {title} {\enquote {\bibinfo {title}
  {Dispersive approximations in fluid dynamics},}\ }\href {\doibase
  10.1002/cpa.3160440102} {\bibfield  {journal} {\bibinfo  {journal} {Comm.
  Pure Appl. Math.}\ }\textbf {\bibinfo {volume} {44}},\ \bibinfo {pages}
  {1--40} (\bibinfo {year} {1991})}\BibitemShut {NoStop}%
\bibitem [{\citenamefont {Majda}\ and\ \citenamefont
  {Timofeyev}(2000)}]{Majda00}%
  \BibitemOpen
  \bibfield  {author} {\bibinfo {author} {\bibfnamefont {A.~J.}\ \bibnamefont
  {Majda}}\ and\ \bibinfo {author} {\bibfnamefont {I.}~\bibnamefont
  {Timofeyev}},\ }\bibfield  {title} {\enquote {\bibinfo {title} {Remarkable
  statistical behavior for truncated {Burgers} {Hopf} dynamics},}\ }\href
  {\doibase 10.1073/pnas.230433997} {\bibfield  {journal} {\bibinfo  {journal}
  {PNAS}\ }\textbf {\bibinfo {volume} {97}},\ \bibinfo {pages} {12413--12417}
  (\bibinfo {year} {2000})}\BibitemShut {NoStop}%
\bibitem [{\citenamefont {Venkataraman}\ and\ \citenamefont
  {Ray}(2017)}]{Venkataraman17}%
  \BibitemOpen
  \bibfield  {author} {\bibinfo {author} {\bibfnamefont {Divya}\ \bibnamefont
  {Venkataraman}}\ and\ \bibinfo {author} {\bibfnamefont {Samriddhi~Sankar}\
  \bibnamefont {Ray}},\ }\bibfield  {title} {\enquote {\bibinfo {title} {The
  onset of thermalization in finite-dimensional equations of hydrodynamics:
  insights from the {Burgers} equation},}\ }\href {\doibase
  10.1098/rspa.2016.0585} {\bibfield  {journal} {\bibinfo  {journal} {Proc. R.
  Soc. A}\ }\textbf {\bibinfo {volume} {473}},\ \bibinfo {pages} {20160585}
  (\bibinfo {year} {2017})}\BibitemShut {NoStop}%
\bibitem [{\citenamefont {Feng}\ \emph {et~al.}(2017)\citenamefont {Feng},
  \citenamefont {Zhang}, \citenamefont {Cao}, \citenamefont {Prants},\ and\
  \citenamefont {Liu}}]{Feng17}%
  \BibitemOpen
  \bibfield  {author} {\bibinfo {author} {\bibfnamefont {Peihua}\ \bibnamefont
  {Feng}}, \bibinfo {author} {\bibfnamefont {Jiazhong}\ \bibnamefont {Zhang}},
  \bibinfo {author} {\bibfnamefont {Shengli}\ \bibnamefont {Cao}}, \bibinfo
  {author} {\bibfnamefont {S.~V.}\ \bibnamefont {Prants}}, \ and\ \bibinfo
  {author} {\bibfnamefont {Yan}\ \bibnamefont {Liu}},\ }\bibfield  {title}
  {\enquote {\bibinfo {title} {Thermalized solution of the {Galerkin}-truncated
  {Burgers} equation: {From} the birth of local structures to
  thermalization},}\ }\href {\doibase 10.1016/j.cnsns.2016.09.004} {\bibfield
  {journal} {\bibinfo  {journal} {Communications in Nonlinear Science and
  Numerical Simulation}\ }\textbf {\bibinfo {volume} {45}},\ \bibinfo {pages}
  {104--116} (\bibinfo {year} {2017})}\BibitemShut {NoStop}%
\bibitem [{\citenamefont {Bec}\ and\ \citenamefont {Khanin}(2007)}]{Bec07}%
  \BibitemOpen
  \bibfield  {author} {\bibinfo {author} {\bibfnamefont {Jérémie}\
  \bibnamefont {Bec}}\ and\ \bibinfo {author} {\bibfnamefont {Konstantin}\
  \bibnamefont {Khanin}},\ }\bibfield  {title} {\enquote {\bibinfo {title}
  {Burgers turbulence},}\ }\href {\doibase 10.1016/j.physrep.2007.04.002}
  {\bibfield  {journal} {\bibinfo  {journal} {Physics Reports}\ }\textbf
  {\bibinfo {volume} {447}},\ \bibinfo {pages} {1--66} (\bibinfo {year}
  {2007})}\BibitemShut {NoStop}%
\bibitem [{\citenamefont {Buzzicotti}\ \emph {et~al.}(2016)\citenamefont
  {Buzzicotti}, \citenamefont {Biferale}, \citenamefont {Frisch},\ and\
  \citenamefont {Ray}}]{Buzzicotti16}%
  \BibitemOpen
  \bibfield  {author} {\bibinfo {author} {\bibfnamefont {Michele}\ \bibnamefont
  {Buzzicotti}}, \bibinfo {author} {\bibfnamefont {Luca}\ \bibnamefont
  {Biferale}}, \bibinfo {author} {\bibfnamefont {Uriel}\ \bibnamefont
  {Frisch}}, \ and\ \bibinfo {author} {\bibfnamefont {Samriddhi~Sankar}\
  \bibnamefont {Ray}},\ }\bibfield  {title} {\enquote {\bibinfo {title}
  {Intermittency in fractal {Fourier} hydrodynamics: {Lessons} from the
  {Burgers} equation},}\ }\href {\doibase 10.1103/PhysRevE.93.033109}
  {\bibfield  {journal} {\bibinfo  {journal} {Phys.\ Rev.\ E}\ }\textbf
  {\bibinfo {volume} {93}},\ \bibinfo {pages} {033109} (\bibinfo {year}
  {2016})}\BibitemShut {NoStop}%
\bibitem [{\citenamefont {Frisch}\ \emph {et~al.}(2003)\citenamefont {Frisch},
  \citenamefont {Matsumoto},\ and\ \citenamefont {Bec}}]{Frisch03}%
  \BibitemOpen
  \bibfield  {author} {\bibinfo {author} {\bibfnamefont {U.}~\bibnamefont
  {Frisch}}, \bibinfo {author} {\bibfnamefont {T.}~\bibnamefont {Matsumoto}}, \
  and\ \bibinfo {author} {\bibfnamefont {J.}~\bibnamefont {Bec}},\ }\bibfield
  {title} {\enquote {\bibinfo {title} {Singularities of {Euler} {Flow}? {Not}
  {Out} of the {Blue}!}}\ }\href {\doibase 10.1023/A:1027308602344} {\bibfield
  {journal} {\bibinfo  {journal} {Journal of Statistical Physics}\ }\textbf
  {\bibinfo {volume} {113}},\ \bibinfo {pages} {761--781} (\bibinfo {year}
  {2003})}\BibitemShut {NoStop}%
\bibitem [{\citenamefont {Gibbon}(2008)}]{Gibbon08}%
  \BibitemOpen
  \bibfield  {author} {\bibinfo {author} {\bibfnamefont {J.~D.}\ \bibnamefont
  {Gibbon}},\ }\bibfield  {title} {\enquote {\bibinfo {title} {The
  three-dimensional {Euler} equations: {Where} do we stand?}}\ }\href {\doibase
  10.1016/j.physd.2007.10.014} {\bibfield  {journal} {\bibinfo  {journal}
  {Physica D: Nonlinear Phenomena}\ }\textbf {\bibinfo {volume} {237}},\
  \bibinfo {pages} {1894--1904} (\bibinfo {year} {2008})}\BibitemShut {NoStop}%
\bibitem [{\citenamefont {Clark~di Leoni}\ \emph {et~al.}(2014)\citenamefont
  {Clark~di Leoni}, \citenamefont {Cobelli}, \citenamefont {Mininni},
  \citenamefont {Dmitruk},\ and\ \citenamefont {Matthaeus}}]{Clark14}%
  \BibitemOpen
  \bibfield  {author} {\bibinfo {author} {\bibfnamefont {P.}~\bibnamefont
  {Clark~di Leoni}}, \bibinfo {author} {\bibfnamefont {P.~J.}\ \bibnamefont
  {Cobelli}}, \bibinfo {author} {\bibfnamefont {P.~D.}\ \bibnamefont
  {Mininni}}, \bibinfo {author} {\bibfnamefont {P.}~\bibnamefont {Dmitruk}}, \
  and\ \bibinfo {author} {\bibfnamefont {W.~H.}\ \bibnamefont {Matthaeus}},\
  }\bibfield  {title} {\enquote {\bibinfo {title} {Quantification of the
  strength of inertial waves in a rotating turbulent flow},}\ }\href@noop {}
  {\bibfield  {journal} {\bibinfo  {journal} {Phys.\ Fluids}\ }\textbf
  {\bibinfo {volume} {26}},\ \bibinfo {pages} {035106} (\bibinfo {year}
  {2014})}\BibitemShut {NoStop}%
\bibitem [{\citenamefont {Clark~di Leoni}\ \emph {et~al.}(2016)\citenamefont
  {Clark~di Leoni}, \citenamefont {Mininni},\ and\ \citenamefont
  {Brachet}}]{Clark16}%
  \BibitemOpen
  \bibfield  {author} {\bibinfo {author} {\bibfnamefont {P.}~\bibnamefont
  {Clark~di Leoni}}, \bibinfo {author} {\bibfnamefont {P.~D.}\ \bibnamefont
  {Mininni}}, \ and\ \bibinfo {author} {\bibfnamefont {M.~E.}\ \bibnamefont
  {Brachet}},\ }\bibfield  {title} {\enquote {\bibinfo {title} {Helicity,
  topology, and kelvin waves in reconnecting quantum knots},}\ }\href@noop {}
  {\bibfield  {journal} {\bibinfo  {journal} {Phys. Rev. A}\ }\textbf {\bibinfo
  {volume} {94}},\ \bibinfo {pages} {043605} (\bibinfo {year}
  {2016})}\BibitemShut {NoStop}%
\end{thebibliography}%

\end{document}